\documentclass[acmsmall]{acmart}

\usepackage{tikz}
\usepackage{amsmath}

\usepackage{booktabs} 
\usepackage{enumerate}
\usepackage{epsfig}
\usepackage{graphicx}
\usepackage{epstopdf}
\usepackage{amsmath}
\usepackage[linesnumbered,ruled]{algorithm2e}
\usepackage{listings}
\usepackage{subfigure}
\usepackage{rotating}
\usepackage{lscape}
\usepackage{verbatim}
\usepackage{hhline}
\usepackage{textcomp,booktabs}
\usepackage{xcolor}
\usepackage{paralist}
\usepackage{grffile}
\usepackage{mathtools}
\usepackage{amsthm}
\usepackage{hyphenat}
\usepackage[autostyle,threshold=0]{csquotes}
\usepackage{tabularx}
\usepackage{CJKutf8}
\usepackage{flushend}

\newcommand{\newmaterial}[1]{\textcolor{black}{#1}}

\def\SurveyTotal{548}
\def\SurveyValid{384}

\setcopyright{acmlicensed}
\acmJournal{PACMHCI}
\acmYear{2025} \acmVolume{9} \acmNumber{N7} \acmArticle{CSCW391} \acmMonth{11}\acmDOI{10.1145/3757572}

\received{October 2024}
\received[revised]{April 2025}
\received[accepted]{August 2025}

\begin{document}

\title[Cooperative Dynamics of Censorship, Misinformation, and Influence Operations]{Cooperative Dynamics of Censorship, Misinformation, and Influence Operations: Insights from the Global South and U.S.}


\author{Zaid Hakami}
\orcid{0000-0003-4236-0349}
\email{zhaka001@fiu.edu}
\affiliation{
  \institution{Florida International University}
  \city{Miami}
  \state{FL}
  \country{USA}
}
\affiliation{
  \institution{Jazan University}
  \city{Jazan}
  \country{Saudi Arabia}
}

\author{Yuzhou Feng}
\orcid{0009-0004-6787-0518}
\email{yfeng015@fiu.edu}
\affiliation{
   \institution{Florida International University}
   \city{Miami}
   \state{FL}
   \country{USA}
}

\author{Bogdan Carbunar}
\orcid{0000-0002-4950-9751}
\email{carbunar@gmail.com}
\affiliation{
   \institution{Florida International University}
   \city{Miami}
   \state{FL}
   \country{USA}
}

\begin{abstract}
Censorship and the distribution of false information, tools used to manipulate what users see and believe, are seemingly at opposite ends of the information access spectrum. Most previous work has examined them in isolation and within individual countries, leaving gaps in our understanding of how these information manipulation tools interact and reinforce each other across diverse societies. In this paper, we study perceptions about the interplay between censorship, false information, and influence operations, gathered through a mixed-methods study consisting of a survey (n = 384) and semi-structured interviews (n = 30) with participants who have experienced these phenomena across diverse countries in both the Global South and Global North, including Bangladesh, China, Cuba, Iran, Venezuela, and the United States. Our findings reveal perceptions of cooperation across various platforms between distinct entities working together to create information cocoons, within which censorship and false information become imperceptible to those affected. Building on study insights, we propose novel platform-level interventions to enhance transparency and help users navigate information manipulation. In addition, we introduce the concept of plausibly deniable social platforms, enabling censored users to provide credible, benign explanations for their activities, protecting them from surveillance and coercion.
\end{abstract}



\maketitle

\section{Introduction}

Censorship activities that control, restrict, or suppress content, and the spread of false information that includes unintentional inaccuracies (misinformation) and deliberate fabrications designed to deceive or manipulate (disinformation), are powerful tools of information manipulation that prevent billions of people worldwide from accessing and interpreting accurate information. These processes are inherently collaborative. Censorship operates through coordination between state-level censors, enterprise-level censors, and ordinary Internet users who engage in self-censorship or report content~\cite{KRC17, KR21, FPSC25, WSH19, W24}. Similarly, the spread of false information depends on cooperation among its creators, consumers, and amplifiers. Influence operations, which often disseminate false information, are by definition collaborative efforts carried out by organized groups pursuing shared agendas and strategic goals~\cite{SAW19, FacebookIO}.

However, these information manipulation tools also have seemingly opposite goals: the distribution of false information introduces undesirable content into public discourse, while censorship restricts access to content deemed sensitive. This partly explains why they have been mostly studied separately, and not in conjunction. Research on censorship has investigated the infrastructure used by censors~\cite{DPRZ08, TAP16, GCNDGSW15, JHWC21, MBVVSSHE18, PJLEFWP17, SSKE20, RSDRSE20, HBSL22}, the development of censorship evasion tools~\cite{HBSL22, BHQL19, KESSMG16, AEYH23} and user perceptions in individual countries in the Global South~\cite{CYL22, GQRW22, WRAR22, LJLNW20, LN21, MDAK17, MAF13, GK17, RVE23, FZSC23, LJLNW20, MAF13, KSN17, WM15} or in the Global North~\cite{DK13, M18, PKG19, RVE23, GS23}. Studies on false information have focused on detection methods~\cite{HALT17, JGBMN18, BML21, NCHAEBPSM21, ZSPLZ22}, user perceptions in specific Global North societies~\cite{SSFS21, KR20, HMWS20, PR19, AMR22, PEMAER21, BRP20} and the Global South~\cite{HYAS20, APKMP21, CP19, APKMP21, SMV22, SF21, WES21}, and interventions aimed at reducing misperceptions~\cite{KWLLMM21, YKBMP20, WCHM20, EORC20}.

In addition, most research has examined perceptions of censorship and false information in individual countries, limiting our understanding of how experiences compare across Global South and Global North societies. Prior work on censorship has also focused on a single type of censor, either state-level~\cite{MAF13, DAB16, GQRW22, LKMRW19, KSN17, WM15}, enterprise-level~\cite{KRC17, M18}, or self-censorship~\cite{DK13, PKG19, LN21, SK11, MAF13, GS23, YL25}, leaving gaps in our knowledge about how people perceive censorship that involves overlapping efforts by states, enterprises, and individual users. More broadly, there remain significant gaps in understanding the complex relationships and collaborative dynamics among various forms of censorship, false information, and influence operations, and the strategies individuals use to navigate or evade their effects, particularly from the perspective of those who experienced these phenomena in the context of diverse countries.

Bridging these gaps is a crucial step for developing a more comprehensive and global understanding of information manipulation, identifying similarities across the tools and contexts in which they are deployed, and revealing both their limitations and new opportunities for designing censorship-resistant systems and defenses against false information that better support users worldwide. To achieve this, we study the interplay between information manipulation tools and examine how they pivot on key issues like content verification, evasion strategies and manipulation platforms, when viewed through the lens of participants who experienced censorship and false information in diverse countries. Specifically, we focus on the following research questions:

\begin{compactitem}

\item
{\it RQ1: How does the impact of diverse types of censorship experienced by Internet users from multiple countries affect their exposure to false information and their efforts to verify content?}

\item
{\it RQ2: How do Internet users from diverse countries experience and perceive the ways in which censorship, false information, and influence operations interact and shape each other?}

\item
{\it RQ3: How do such users navigate and evade environments controlled by information manipulation tools?}

\end{compactitem}

\noindent
To investigate these questions, we conducted a mixed-methods study consisting of a survey ($n$ = \SurveyValid) and semi-structured interviews ($n$ = 30) with users who have experienced censorship and false information in diverse Global South countries and the United States. The study examines experiences with censorship, its impact and the countries involved, and also experiences with false information and influence operations, and perceptions of the interplay between these information manipulation tools.

We identify complex relationships, including that censorship and false information, and also censorship and influence operations, mutually enable, amplify and even justify each other. Survey respondents who experience higher-impact censorship tend to encounter false information more frequently and tend to evade censorship more frequently to verify information. However, more than 79\% of respondents agree or strongly agree that censorship makes information verification more difficult (Section~\ref{sec:survey}). Interviews reveal that influence operatives and even regular users, exploit censorship to distribute false information. Conversely, false information and influence operations are also used to justify censorship. Influence operations use social media both to spread false information and to reinforce censorship by suppressing dissent and undermining information verification strategies (Section~\ref{sec:interviews}).

Our investigations into the interplay among censorship, false information, and influence operations, and the strategies users employ to navigate them, offer insights for CSCW research on socio-technical cooperation and resistance. We build on prior CSCW work on online collaboration in censored environments~\cite{CYL22} and coordinated influence operations~\cite{SAW19} by theorizing the cooperative dynamics of information control. Our study reveals how distinct actors and institutions (e.g., governments, media platforms, educational systems) coordinate censorship and false information across multiple infrastructures to construct {\it information cocoons}, i.e., closed informational environments where manipulation becomes difficult to perceive. This re-frames information cocoons from emergent side effects of personalization~\cite{S01, P11} to strategically engineered outcomes.

We also contribute to the CSCW concept of {\it resilience}, the capacity to adapt to disruption through new social or technological routines~\cite{KW03, MS08, MAS09, VD17, S19, PKVZ23}. In line with this literature, our findings show how individuals and communities in censored contexts develop verification routines and evasion strategies, often relying on strong-tie networks, coded language, or border-crossing technologies, to circumvent information control and restore epistemic agency (Section~\ref{sec:discussion:design}). These evasion strategies are not only signs of coping, they show that trust and information practices depend on people’s contexts and relationships, especially under information cocoons.

In addition, we extend CSCW and HCI frameworks on {\it appropriation} and {\it counter-appropriation}~\cite{WRAR22, LKMRW19, SP18} by revealing how state actors appropriate participatory infrastructures like social media for control and propaganda and how users re-purpose technologies to subvert both censorship and false information. We further show how these appropriations are actively contested, revealing a co-evolution of control and resistance strategies.

We distill study findings into recommendations for designing social platforms to help users navigate information manipulation within information cocoons (Section~\ref{sec:discussion:design}). We leverage these recommendations to propose novel platform interventions aimed at increasing awareness of the interplay between censorship, false information, and influence operations, by providing users with more transparency, context, and options regarding the content they access (Section~\ref{sec:design:interventions}). In addition, we propose an approach to allow users to present upon coercion from authorities, a plausible and benign explanation for social media activities performed while evading censorship (Section~\ref{sec:design:plausible}). While this approach was originally intended to address participant-reported needs for private and plausibly deniable censorship evasion in countries like Iran and China, more recent events also suggest its relevance for people traveling to the U.S.~\cite{PhoneInspections, USNPhone}.

\section{Background and Related Work}
\label{sec:related}

\subsection{Perspectives on Internet Censorship and Circumvention}

Internet censorship that includes blocking, filtering, surveillance, and content removal, is carried out by state, enterprise, and user actors~\cite{DPRZ08, TAP16, GCNDGSW15, JHWC21, KRC17, KR21, XK19, VP15, LZSGLGG22}. Censorship is a layered, decentralized process shaped by political agendas and commercial interests. While states  use censorship to consolidate power and assert infrastructural control, they often justify it through appeals to national security or morality~\cite{D13}. Enterprises, particularly social and communication platforms, enforce censorship under a combination of state pressure~\cite{MetaRemoval}, legal mandates~\cite{EUDSA, GDPRChildren, COPPA}, and internal policies~\cite{FacebookMisinformation, TwitterMisinformation}. These platforms frequently need to navigate tensions between authoritarian demands and democratic values. Censorship is also participatory, where users report violations~\cite{WSH19, W24, InstagramReport, XReport}, engage in self-censorship, and police others through cancel culture. This ecosystem blurs the line between public and private authority, making censorship a distributed and negotiated practice~\cite{KRC17, KR21, FPSC25, WSH19, W24}.

\textbf{State-level censorship} is implemented through diverse strategies that reflect differences in political priorities and technological capacities. In countries like China and Iran, censorship combines legal frameworks, ideological control, and technical infrastructure to enforce a tightly regulated information environment~\cite{WSSBAWBHLW23, HNDKLMCNGP21, BNRL21, A13, AAH13, BFRSL20}. In Bangladesh, by contrast, authorities often employ network shutdowns and selective content restrictions during periods of political unrest.
Further comparative research in Global South countries reveals that, while the Internet was initially considered a tool for promoting liberalization~\cite{H11, MAF13, DAB16}, increasing state control has shifted it toward a mechanism of political consolidation. {\it Internet nationalization}~\cite{GQRW22} is a concept used to describe this evolution, where governments not only regulate content but also assert ownership over infrastructure, impose data localization, and restrict foreign services. Such infrastructural control suggests that censorship perpetuates {\it infrastructural violence}, by embedding repression into the architecture of Internet access and use~\cite{RO12}.

Studies of {\bf enterprise-level censorship} implementation have mainly focused on social media and communication platforms in China, revealing that platforms primarily use keyword filtering to censor content~\cite{KRC17, KR21, XK19, VP15}. Despite state influence, platforms retain some autonomy in defining censored keywords~\cite{KRC17}, indicating a complex interplay between government mandates and corporate discretion. Studies of user perceptions of platform censorship have examined however countries from both the Global South and Global North, revealing regional differences and common ambiguities. In China, while attitudes toward censorship include acceptance, criticism, and indifference~\cite{LJLNW20}, it is unclear if the study's users distinguish state-imposed censorship from platform-level content moderation. In contrast, in Weibo, Chinese LGBT+ communities engage more explicitly with enterprise censorship, e.g., by identifying and navigating the platform's filtering of LGBT+ content~\cite{CYL22}. Similar perceptions appear in Global North contexts~\cite{M18}. For instance, U.S. participants often view content moderation to be censorship that not only removes speech but also restricts future expression (e.g., account suspensions). Further, they sometimes suspect platforms of attempting to influence elections~\cite{M18}. This reveals that across contexts, enterprise-level censorship is intertwined with political and social power struggles even though it is operationally distinct from state censorship. This questions simple distinctions between state versus corporate control.

{\bf Self-censorship}, both a form of censorship and a strategy to evade external restrictions, shapes online communications across both Global South and Global North societies~\cite{LN21, SK11, MAF13, YL25, DK13, PKG19, GS23}. Self-censorship is not only an individual coping mechanism, but also a channel through which political control and social pressure operate. In China, early studies suggested that censorship did not substantially induce self-censorship~\cite{R15}. However, recent work has re-framed it to be a ``micro-foundational'' mechanism that spreads authoritarianism through social influence without explicit coercion~\cite{YL25}. Furthermore, also in China, self-censorship has been linked to political disengagement and discourse avoidance due to fear of surveillance or criticism~\cite{MAF13}.

These behaviors and motivations are not unique to authoritarian regimes. In the U.S. and UK, many Facebook users were shown to engage in last-minute self-censorship~\cite{DK13}, while exposure to disagreeable views on social media discourages opinion expression~\cite{GZ15}. During the 2016 U.S. elections, students preferred offline political discussions to avoid online conflict~\cite{PKG19}. In the U.S., similar to China, people often self-censor because they feel divided from those with opposing political views or due to perceptions of government repression~\cite{GS23}. These parallels suggest that despite structural differences, the psychological and social dynamics of self-censorship operate similarly across regime types. This also supports the {\it spiral of silence} theory~\cite{NN74}, where fear of isolation or backlash drives users to withhold minority views, even in open societies. This suggests that both censorship regimes and social conformity can suppress discourse.

\noindent
{\bf The Cyclical Evolution of Censorship and Circumvention}.
Censorship disproportionately affects resource-constrained users in the Global South, increasing digital inequalities between those who can circumvent restrictions and those who cannot~\cite{MDAK17, DAB16}. These constraints shape both online and offline behavior~\cite{MDAK17, OLZ11, WAAASRYR13, MAS09}, and often prompt users to develop creative workarounds~\cite{MDAK17, DAB16}. This dynamic is not static: censorship triggers a recursive cycle of appropriation and counter-appropriation~\cite{LKMRW19}, where state actors upgrade control mechanisms, and users respond with evasive tactics, especially among younger generations~\cite{WRAR22}. For example, recent gaming restrictions for minors in China have unintentionally escalated censorship evasion behaviors~\cite{FZSC23}. However, flawed mental models about the efficacy and risks of circumvention tools can increase user exposure and provide new surveillance vectors~\cite{RVE23}.

Attitudes toward censorship and its evasion can be ambivalent and contradictory~\cite{KSN17, WM15}. In China, even users skilled at bypassing censorship and who share sensitive or restricted information, sometimes accept restrictions in specific contexts~\cite{KSN17}. Support for censorship also varies, with individuals scoring high on authoritarian personality measures being more likely to endorse it~\cite{WM15}.

\noindent
{\bf Research Gaps}.
Existing literature often considers discrete censorship categories (state, enterprise, or user-driven), limiting understanding of how people experience censorship that involves multi-layered socio-technical systems. This fragmentation overlooks how these actors interact to control information. Moreover, studies tend to focus on single country cases, often in the Global South. This leaves gaps in comparative insights across political and cultural contexts, and constrains our ability to theorize how infrastructure, power, and user agency shape information access. Our research questions address these gaps by examining how diverse forms of censorship affect user exposure to and verification of false information (RQ1), how users perceive the entanglement of censorship, misinformation, and influence operations (RQ2), and how they navigate and resist these intertwined systems (RQ3), across both Global South and North contexts.

\subsection{Perspectives on False Information and Influence Operations}

The term false information encompasses both misinformation and disinformation. Misinformation refers to false or inaccurate information that may be spread unintentionally or without malicious intent, while disinformation refers to false information that is intentionally fabricated to deceive or influence an audience, often for political, social, or economic gain. Influence operations (IOs) are coordinated efforts to distribute false information, propaganda, and selectively framed content to promote a specific agenda and shape public perceptions, opinions, or behaviors toward a strategic goal~\cite{FacebookIO}.

\noindent
{\bf Perceptions of False Information in Global North and Global South Contexts}.
Research on false information reveals both contextual differences and shared challenges in how users, journalists, and fact-checkers evaluate and respond to false content across Global North and South settings~\cite{SSFS21, KR20, HMWS20, PR19, AMR22, PEMAER21, BRP20, HYAS20, APKMP21, CP19, SMV22, SF21, WES21, SPLGBV24}. Across regions, user judgments are shaped by a combination of cognitive, social, and emotional factors. However, their relative influence varies. In the Global North, belief alignment often shapes user judgments, though factual accuracy can still carry more weight~\cite{PR19, AMR22, PEMAER21, BRP20}. Visual content often outweighs source credibility in shaping perceptions~\cite{SSFS21}, revealing the persuasive power of images across platforms. Perceived accuracy, however, has limited influence on sharing behavior~\cite{PEMAER21, PMZLR20}. This suggests motivations beyond truth evaluation, e.g., identity signaling or emotional expression~\cite{PEMAER21, PMZLR20}.

These challenges are further magnified in the Global South, due to structural inequalities and misinformation exposure and response. While journalists in the Global North often rely on expert consultation and manual verification instead of computational tools~\cite{HMWS20}, such practices are less feasible in resource-constrained contexts. In Bangladesh, fact-checking is not widely considered to be part of journalistic responsibility, despite public expectations~\cite{HYAS20}. In rural India, the circulation of falsehoods during the COVID-19 pandemic was not merely due to information scarcity but also fueled by affective polarization and communal divisions~\cite{APKMP21, SPLGBV24}.

These findings suggest a critical asymmetry. While false information abounds in both regions, journalists, fact-checkers, and users in the Global South bear a heavier verification burden due to limited institutional support, weak technological infrastructure, and lower epistemic trust. This imbalance challenges the universality of many misinformation interventions, often designed for and tested in Global North settings. This further questions the ability of existing solutions to address the vulnerabilities and different media norms of the Global South~\cite{HYAS20, APKMP21, SPLGBV24}, and suggests the need to develop approaches grounded in local information ecosystems and power structures.

\noindent
{\bf Perceptions of Influence Operations}.
Starbird et al.~\cite{SAW19} first argued for studying influence operations (IOs) within CSCW, emphasizing their collaborative and distributed nature, features central for this research community. Since then, IOs have been documented globally~\cite{FacebookIO}, though their mechanisms, targets, and political stakes differ across contexts. In the Global North, research and media attention has focused on foreign state actors, particularly Russia and China, whose IOs aim to disrupt democratic processes and shape geopolitical narratives~\cite{IRA, NYT.Trolls, H15, M17}. These campaigns, often run by centralized {\it troll armies}~\cite{IRA, NYT.Trolls, KPR17}, have created a blueprint that is now used in Global South countries, where domestic actors deploy similar techniques in pursuit of internal political control. Governments in Venezuela~\cite{Venezuela.Trolls}, Iran~\cite{ZCSSSB19}, Turkey~\cite{B21}, and India~\cite{DM24} have adopted IOs to suppress dissent, manufacture consensus, and control narratives~\cite{H15, DM24}.

While this suggests increasingly similar tactics worldwide, IOs still operate within distinct political contexts and communication goals. In the Global North, the dominant frame is defensive, focused on protecting electoral integrity from foreign interference. In contrast, IOs in the Global South often reflect proactive state strategies to shape domestic opinion, frequently using nationalism to mobilize online workers~\cite{DM24, B21}. This points to contextual asymmetries in who performs IOs, what motivates them, and how they are framed and contested.

With improvements in detection techniques~\cite{KSSY20, SKSY22, VET20, LPBE24}, recent research has turned toward understanding how IOs are experienced and resisted by different actors. Hacktivists, for example, consider IOs to be threats to democratic and participatory ideals of the Internet, and engage in countermeasures that include doxxing operatives and mapping their recruitment and funding networks~\cite{SK23}. In contrast, interviews with operatives in Venezuela provide rare insider views of how they adapt to platform defenses and deploy evasion strategies to sustain influence~\cite{RCHS23}. These studies reveal a complex interplay between influence, resistance, and platform governance, shaped by both global platform policies and local power structures~\cite{SK23, SDPJ23, RCHS23}.

\noindent
{\bf Research Gaps}.
Previous research has largely examined user perceptions of false information in either Global North or Global South contexts, and typically considers false information, influence operations, and censorship to be separate domains. This fragmentation limits CSCW's understanding of how users experience the combined effects of information manipulation across diverse socio-political contexts (RQ1, RQ2). In addition, while studies have explored verification and avoidance behaviors, less is known about how these practices function like adaptive strategies within contested information environments. This raises a critical open question: how do users across different contexts navigate and resist information manipulation infrastructures? (RQ3). Addressing this gap contributes to CSCW concepts of user agency, resilience, and the governance of digital platforms.

\subsection{Relationship Between Censorship and False Information}

Two recent studies have examined relationships between censorship and false information online. First, while early research found that China's censored media environment can encourage more critical news consumption~\cite{MAF13}, a more recent study explored how censorship and astroturfing influence the perceptions of WeChat users about information credibility~\cite{LJLNW20}, revealing that confusion over online comments leads many to favor censorship. The authors call this the {\it government’s dividend}, where low-quality information, often directed by the government, can reinforce public support for interventions and increase the perceived need for censorship. A second study offers an ethnographic account of infrastructural violence~\cite{RO12} experienced by Chinese citizens during the early months of the COVID-19 pandemic, documents the epistemic, emotional, and interpersonal harms produced by censorship and propaganda, and reveals how they restrict access to truthful information and disrupt personal relationships~\cite{LN21}.

\noindent
{\bf Research Gaps}.
While these studies provide valuable insights into how censorship and state-aligned information shape public perception, they are limited in scope and integration. They focus exclusively on China. They examine individual censorship, false information, or influence operations mechanisms in isolation, but not their coordinated interplay. Even when relationships between these mechanisms are noted, e.g., in the concept of the government’s dividend~\cite{LJLNW20}, the studies do not treat their interaction to be a core analytic focus and do not explore how these forces operate together to shape user experience. Moreover, they do not examine how perceptions and adaptation strategies vary across different socio-political contexts.

Our study addresses these gaps by investigating how users from diverse countries perceive the interplay between censorship, false information, and influence operations. We examine how these mechanisms shape both information access and user agency, and how users adapt. We employ a cross-national, mixed-methods approach that broadens existing perspectives and offers a more integrated view of how information control operates across contexts.

\subsection{Information Cocoons}

The concept of {\it information cocoons} has been used to describe environments where individuals are shielded from diverse perspectives, often reinforcing their beliefs. Sunstein’s notion of information cocoons~\cite{S01} and Pariser’s filter bubble~\cite{P11} focus on the role of self-selection and algorithmic personalization in shaping exposure to information online. These concepts have been expanded in studies of echo chambers, where social media users tend to cluster in ideologically homogeneous communities that amplify false information and obstruct critical evaluation~\cite{CDGQS21, TGBVSSSN18}. These studies help explain how restricted or biased exposure to information may limit the ability of users to verify content, thus are related to RQ1. However, they focus on user or algorithm-driven dynamics within relatively open media environments, where access to alternative views is theoretically available.

In contrast, research on state-sponsored information control has focused on systemic manipulation instead of individual behavior. Studies of China’s censorship and propaganda reveal how the state combines suppression, distraction, and false information to shape public opinion and prevent dissent~\cite{KPR13, KPR17}. However, this work overlooks how censorship is implemented through a multi-layered system that involves not only states but also enterprises and users. Our study shifts the focus from state-level strategies to how users perceive and navigate the interplay of censorship, misinformation, and influence operations across these layers. This perspective captures experiences of information control, and informs RQ2, exploring how users interpret these coordinated dynamics.

The concept of {\it infrastructural violence}~\cite{RO12} further frames these systems to be structural barriers to reliable knowledge. While prior work has examined evasion strategies in single country contexts, there is little research on how these strategies vary across diverse political and technological settings, and how they interplay with censorship and false information campaigns. This gap is central to RQ3, which examines evasion and verification practices of information cocoons across diverse national contexts.

\section{Methods}
\label{sec:methods}

This paper presents results from a mixed-methods approach that combines a survey with in-depth interviews for qualitative insights.

\subsection{Survey Study}
\label{sec:methods:survey}

We designed an online survey to explore how the impact of censorship affects exposure to and verification of false information (RQ1) and to help us recruit diverse interview participants, see Appendix~\ref{appendix:survey}. The survey introduction defines false information to be content that is inaccurate or misleading, whether intentional (e.g., propaganda, fabricated or manipulated content) or unintentional (e.g., factual mistakes, unverified claims), and censorship to be the control or suppression of online information, which may involve blocking access to certain websites, filtering or removing content, and restricting the ability of individuals to communicate freely online. The introduction also provides examples of services that can be censored or be engaged in censorship, including social networks, communication apps, search engines, news sites, cloud storage services, online payment services, cryptocurrencies, and VPN providers.

\noindent
{\bf Measures}.
To encourage completion without compensation, the survey was limited to ten items. The first item identifies countries where participants experienced censorship and respondents who did not experience censorship. Subsequently, the survey examines several constructs. For participants who experienced censorship, the {\it impact of censorship} construct is investigated through two items that assess the perceived impact of the most significant censorship they encountered in the U.S. and (separately) abroad, on a scale from 1 (no impact at all) to 5 (extreme impact).  The {\it impact of censorship on false information} construct is evaluated through two Likert-scale items, where one measures perceptions of censorship hindering verifying false information that circulates online (for all respondents), while the other investigates how often censored individuals have evaded restrictions to verify information. The {\it frequency of encountering false information online} construct is measured by two Likert-scale items: one for respondents without censorship experience and another for those who faced censorship. Both items asses how frequently the participants encountered false information online.

\noindent
{\bf Survey Validation}.
We conducted pre-tests with 21 participants (16 students and 5 faculty, including experts) who were exposed to censorship and false information in Bangladesh, China, Cuba, India, Iran, Saudi Arabia, Turkey, the U.S., and Venezuela. Data from the pre-tests is not included in our final results. Participant responses and feedback were instrumental to enhance the clarity and neutrality of the survey questions, by helping us identify and remove ambiguities, misunderstandings, and biases. For example, we replaced initial use of the word `fact-check' with `verify', and, for the items that assess the perceived impact of the most significant censorship, we included response options for participants who have not experienced censorship in either context. In addition, we removed items assessing the services where participants experienced censorship, which we moved to the survey introduction.

The survey defined censorship broadly to include blocking access, content removal, and communication restrictions. However, it did not distinguish between types of censors (e.g., state vs. enterprise). However, pre-test feedback revealed that participants in the U.S. often interpreted censorship to be enterprise or platform-based, while participants in Global South countries described state-imposed restrictions. This interpretive variability was retained to reflect participants’ lived experiences but should be considered when comparing responses across countries.

\subsection{Interview Study}

Semi-structured interviews were conducted to explore participant experiences with censorship, false information, and influence operations across countries they lived in, see interview guide in Appendix~\ref{appendix:interview}. Topics included censorship types, censors, evasion attempts, fact-checking behaviors, and exposure to influence operations. Interviews also investigated links between censorship, false information, influence operations, and information verification behaviors. To reduce social desirability bias, several questions were framed around instances experienced by the participants.

Interviews with local participants were conducted in person. Interviews with remote participants took place via Zoom, WhatsApp, or Skype. All interviews were conducted in English. Interviews started with the consenting process, including a description of the team and the study objectives, procedures, and risks. Audio was recorded with consent. Interviews lasted about an hour (M = 58.93 mins, SD = 15.61 mins).

Interviews were transcribed, pseudonymized, and securely stored. Participants were assigned IDs based on participation order. Transcripts were analyzed using applied thematic analysis~\cite{GME11, GME12}, by systematically generating and iteratively conceptualizing codes and themes. Two researchers independently coded the first five transcripts using an initial codebook, then met to compare and discuss coding decisions. Disagreements were resolved through collaborative dialogue, which clarified interpretations, refined code definitions, and updated the codebook. When consensus could not be reached during these meetings, a more senior team member was consulted to help adjudicate and finalize decisions. This process was repeated in batches of 3–5 transcripts, with researchers meeting after each batch to reconcile differences and reach consensus on emerging themes. After finalizing the codebook, both researchers independently recoded all transcripts, followed by a final round of comparison and discussion to ensure consistent application. Instead of relying on formal intercoder reliability statistics, we prioritized interpretive alignment and reflexivity~\cite{MSF19}, addressing disagreements through iterative consensus-building to enhance the rigor and credibility of the thematic structure. Participants received 30 USD.

\noindent
{\bf Cross-National Experience Analysis}.
Several participants had lived in multiple countries with varying levels of information control. During interviews, we asked participants to specify which country they were referring to when discussing their experiences. In the thematic analysis, we annotated excerpts with country context, then compared themes across countries to identify context-specific and cross-cutting patterns. This allowed us to examine how transnational experiences informed user understanding of manipulation strategies and shaped their responses.

\subsection{Participant Recruitment}

We recruited study participants over a seven-month period, from March to October 2024. We aimed to identify participants who experienced false information and censorship with diverse impacts across various countries. Recruitment took place at a U.S. university via emails to local student groups, WhatsApp groups of international students and local student organizations, and posters displayed at various locations across the campus, including the library and International Student and Scholar Services office. The emails included the survey link, while posters and social media posts contained a QR code for the link.

At the end of the survey, participants could express interest in a one-on-one interview and provide contact details. To ensure diversity, invitation preference was given to respondents based on the countries in which they had experienced censorship and its impacts. The recruitment process further used snowball sampling~\cite{BW81}, where each interview participant was asked to share the survey link with their contacts. This approach facilitated access to, and helped build trust with, individuals from diverse backgrounds within a hard-to-reach population who had experienced censorship and were willing to share their experiences. This ensured that participants in both the survey and interview were recruited from diverse locations and had experienced censorship across diverse countries.

\begin{table}
\centering
\resizebox{0.59\columnwidth}{!}{
\textsf{
\begin{tabular}{llcc} 
\toprule
Demographic & Group & \# & \% \\
\midrule
Gender & Male   & 207    & 53.91\%        \\
       & Female & 122  & 31.77\%      \\
       & Non-Binary & 8   & 2.08\%   \\
       & Prefer not to answer   & 47   & 12.24\%  \\
\midrule
Age & 18--25 & 205 & 53.39\% \\
    & 26--35 & 109 & 28.39\% \\
    & 36--45 & 16 & 4.17\% \\
    & 46--55 & 8 & 2.08\% \\
    & Above 56 & 5 & 1.3\% \\
    & Prefer not to answer & 41 & 10.68\% \\
\midrule
Censorship & Country    & \# Respondents & Internet Freedom \\
           &            &    & Score (IFS)~\cite{FREEDOMNET}\\
                            & USA &  192 & 76 \\
                            & Cuba &  33 & 20 \\
                            & China &  29 & 9 \\
                            & Bangladesh &  22 & 41 \\
                            & India &  16 & 50 \\
                            & Venezuela &  15 & 29 \\
                            & Iran &  13 & 11 \\
                            & Turkey &  11 & 30 \\
                            & Pakistan &  6 & 26 \\
                            & Saudi Arabia &  5 & 25 \\
                            & United Arab Emirates &  5 & 30 \\
                            & Colombia &  4 & 65 \\
                            & Russia &  4 & 21 \\
\bottomrule
\end{tabular}}}
\caption{Demographics of survey respondents.}
%
%
\vspace{-20pt}
\label{tab:survey:demographics}
\end{table}

\subsection{Ethical Considerations}

Participants were informed of potential risks, including emotional discomfort from discussing past experiences with censorship and evasion activities, and had the option to skip questions. Risks were mitigated for some participants by conducting the interviews outside restrictive jurisdictions. The procedure was scrutinized and approved by the institutional review board at our university.

During the recruitment process, researchers had no access to personally identifiable information (PII) of users. The survey collected only necessary contact details from respondents interested in interviews, and excluded real names or other identifiers. After processing, all contact and payment data were discarded, to prevent linking participants to responses. Contact information of those not interviewed was destroyed. No participant PII was stored during payment. Interview responses were pseudonymized, stored on a secure university Linux server, and accessed only via encrypted channels from the authors' password-protected laptops.

\subsection{Study Limitations}

Survey respondents represented 31 countries. Not all countries with documented censorship were included, and only six countries were represented by at least ten respondents. Our findings cannot be generalized to countries from which no participants were recruited. Moreover, since most participants were students recruited from a university in the U.S., who are likely more educated and potentially from more affluent backgrounds, our findings may not be generalizable to the broader populations of the represented countries.

Participant responses may be influenced by self-reporting bias, e.g., overreporting, underreporting, and social desirability. To help mitigate these effects, participants were asked to provide concrete examples from their own experiences. Our recruitment methods may have introduced self-selection bias because most (but not all) participants were living in the U.S. during the study. Expatriates may perceive censorship in their home countries differently, and potentially more critically, than current residents. Snowball sampling may have contributed to increased participant homogeneity: people are more likely to recruit others with similar backgrounds and experiences.

The survey does not fully distinguish between different types of censorship (e.g., state, enterprise, user-level), which may have shaped how participants interpreted questions. While the survey introduction defines censorship broadly, pre-test findings indicated that interpretations varied by context, especially between U.S. and Global South respondents. We retained this interpretive flexibility to reflect the experiences of participants, but acknowledge it to be a limitation for cross-national comparison. Further research should examine in more depth the impact of different censorship sources.

While several items in our survey measure the same concept they are not expected to produce similar results, thus we have not evaluated their internal consistency reliability. We also did not validate participant responses against behavioral measures. Such validation is difficult for participants who experienced censorship in other countries in the past, and we also sought to minimize risks for participants who, at the time of the study, were actively experiencing censorship in countries like China and Iran.

\section{Survey Results}
\label{sec:survey}

The survey received responses from \SurveyTotal\ participants over 18 years old. We discarded data from participants who did not consent, did not answer any question, provided inconsistent answers, or did not experience censorship. Table~\ref{tab:survey:demographics} shows demographics of the remaining \SurveyValid\ survey respondents. Most respondents are undergraduate and graduate students, but also include educators, artists, engineers, healthcare and IT professionals. The survey was completed from 15 different countries including the U.S., Bangladesh, China, Saudi Arabia, Venezuela, and Russia.

Bonferroni correction was applied to statistical tests to adjust significance levels and reduce the risk of Type I errors.

\begin{figure*}
\centering
\subfigure[]
{\label{fig:country:impact}{\includegraphics[width=0.49\textwidth]{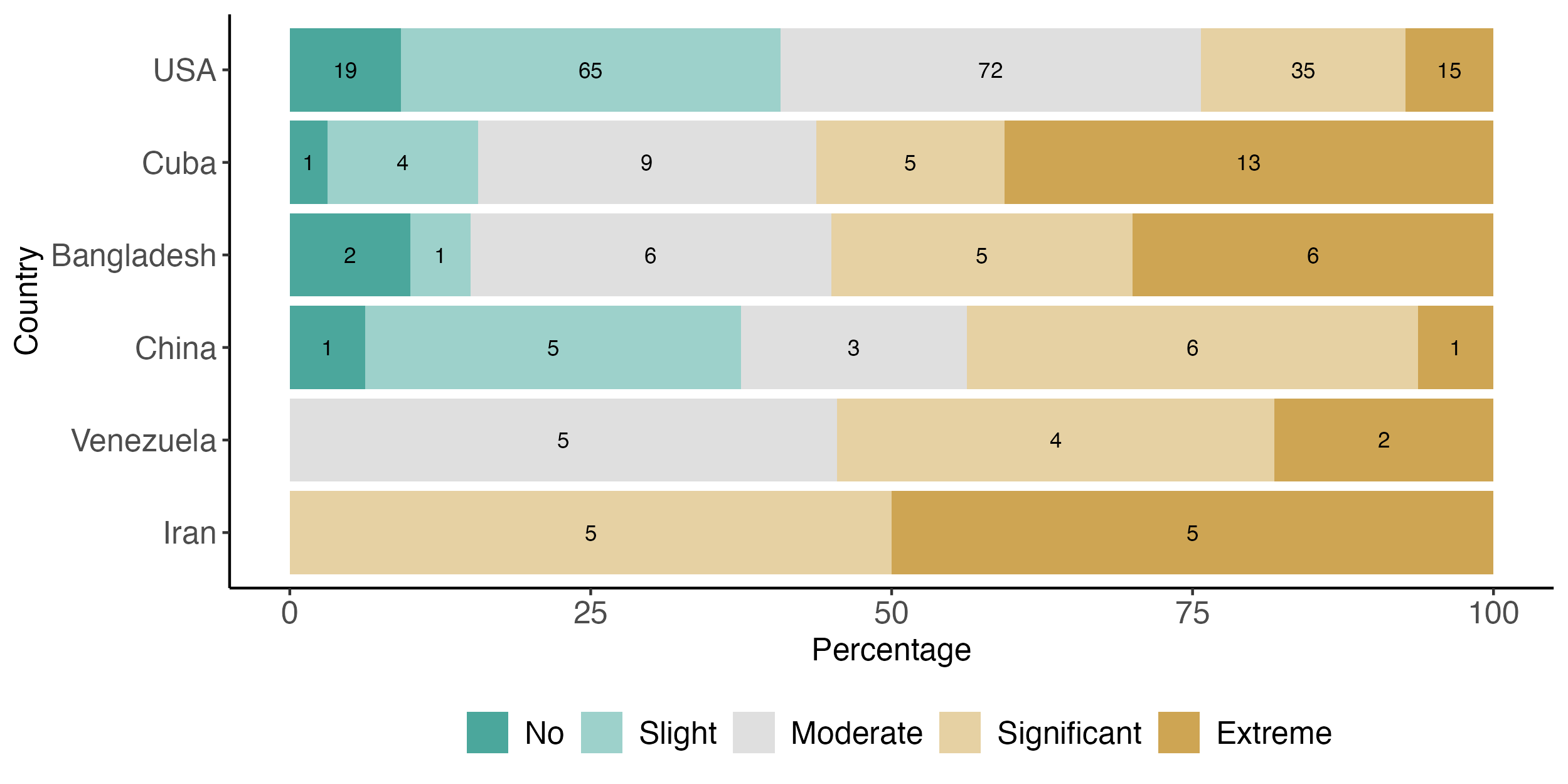}}}
\subfigure[]
{\label{fig:falseinfo}{\includegraphics[width=0.49\textwidth]{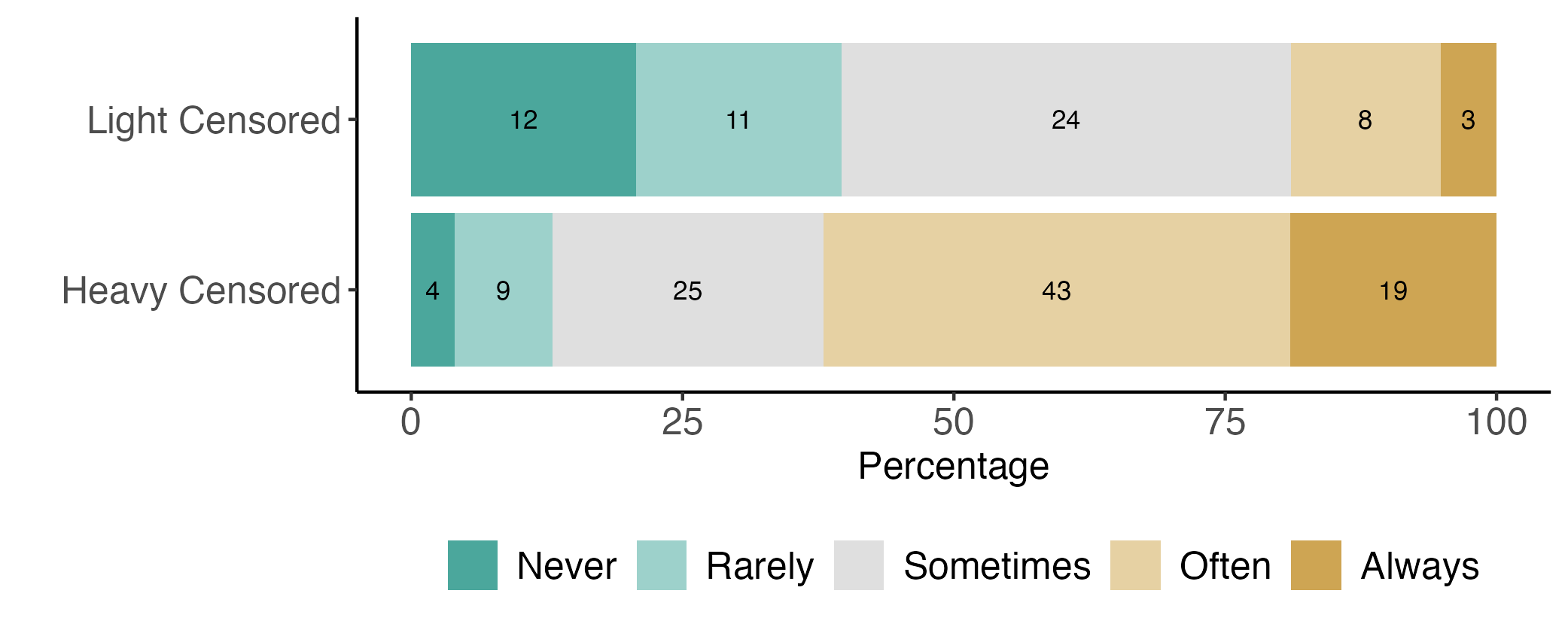}}}
\vspace{-10pt}
\caption{(a) Per-country distribution of participant perception of impact of censorship. All the survey respondents who experienced censorship in Iran and Venezuela perceived it had at least moderate impact. However, for several participants from U.S., Cuba, and China, the impact was inconsistent with the country's IFS score, see Table~\ref{tab:survey:demographics}.
(b) Distribution of frequency of exposure to false information. A higher percentage of highly censored respondents encountered false information often or always when compared to lightly censored ones. Differences are statistically significant.}
\vspace{-15pt}
\end{figure*}

\noindent
{\bf Countries and Impact of Censorship}.
Figure~\ref{fig:country:impact} shows the per-country distribution of perceptions of the worst impact of censorship experienced by survey respondents. The survey did not differentiate between types of censors, see also Section~\ref{sec:methods:survey}. Interpretations of ``censorship'' may thus reflect state-imposed restrictions in some countries and enterprise or platform-level moderation in others. This contextual difference is particularly relevant when comparing perceptions of the impact of censorship across countries and is further examined in the interview findings.

To investigate whether there is an association between perceptions of the impact of the most significant censorship experienced, and the countries where it occurred, we used Chi-square tests of independence, to tell us if there are differences between the impact of the worst censorship experienced in various countries. Pairwise tests did not reject the null hypothesis for Bangladesh, Cuba, China, and Venezuela. This indicates that there is no statistically significant evidence of association between the perceived impact of censorship and these countries, thus, no statistically significant differences in the perceived impact of the worst censorship experienced in these countries.

A Chi-square test of independence investigating this relationship for the U.S. and China also did not reject the null hypothesis (\(\chi^2\) = 4.7, p-value = 0.31), indicating no statistically significant differences in the perceived impact of the worst censorship experienced in these countries. However, caution is warranted in interpreting this result. This is because perception of censorship may stem from different underlying mechanisms, for instance, corporate content moderation in the U.S. versus state-imposed Internet controls in China.

Figure~\ref{fig:country:impact} shows that even in the U.S., a country ranked `Free' by Freedom House~\cite{FREEDOMNET}, 23.86\% (42 out of 176) respondents reported experiencing censorship with at least a significant impact; for a third of those who experienced censorship in China, a country ranked `Not Free'~\cite{ChinaFreedom}, the impact was at most slight. In the next section, we provide interview insights that help explain this finding.

However, Chi-square independence tests revealed a statistically significant association between perceptions of the impact of censorship and the country where it was experienced in the comparisons of U.S. vs. Cuba (\(\chi^2\) = 31.28, p-value = $2.69 \times 10^{-6}$) and the U.S. vs. Iran (\(\chi^2\) = 31.98, p-value = $1.93 \times 10^{-6}$). This suggests significant differences in how respondents from these country pairs perceive the impact of censorship, see Figure~\ref{fig:country:impact}: 51.42\% of respondents from Cuba and 100\% from Iran experienced significant or extreme impact of censorship, compared to 24.27\% in the U.S.

\begin{figure*}
\centering
\subfigure[]
{\label{fig:evade}{\includegraphics[width=0.49\textwidth]{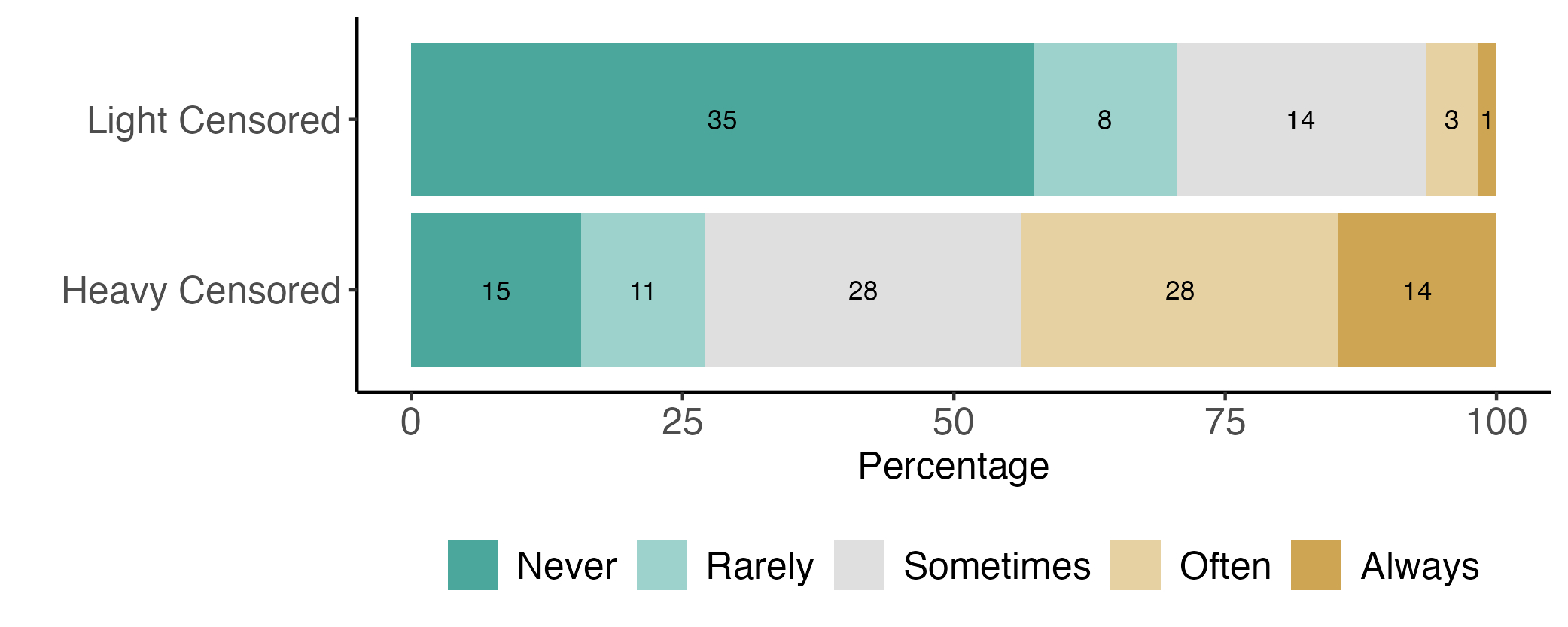}}}
\subfigure[]
{\label{fig:verify}{\includegraphics[width=0.49\textwidth]{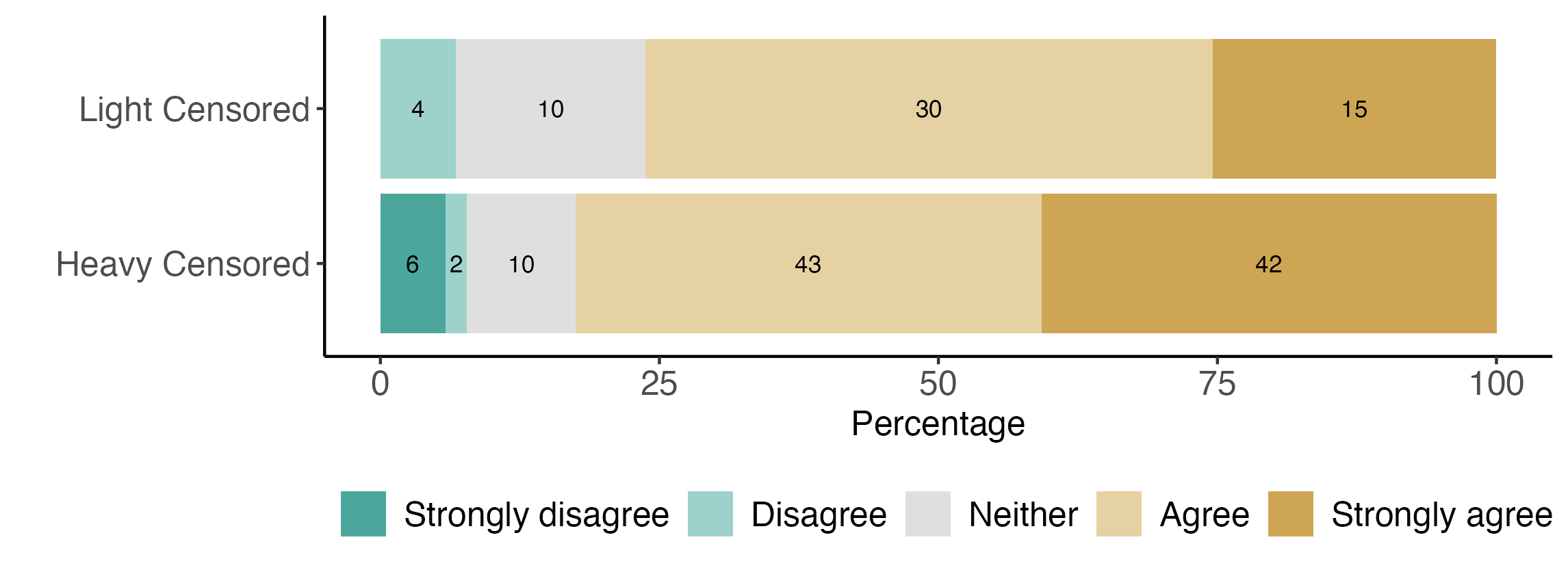}}}
\vspace{-10pt}
\caption{(a) Distribution of frequency of participants evading censorship to verify information. Almost half of heavily censored participants evaded censorship often or always to verify information, while almost 3/4 of lightly censored ones have never or rarely done it. (b) More than 75\% of respondents agree or strongly agree that censorship makes it harder to verify false information. The percentage is higher for respondents who experienced heavy censorship.}
\vspace{-15pt}
\end{figure*}

\noindent
{\bf Exposure to False Information}.
In the following, we say a participant experienced {\it light}  censorship if censorship had no or slight impact, and {\it heavy} censorship if the impact was significant or extreme. Figure~\ref{fig:falseinfo} compares the distribution of the frequency of encountering false information for respondents who experienced light vs. heavy censorship. Participants experiencing heavy censorship were more likely to encounter false information often or always (62\%) compared to those experiencing light censorship (18.97\%); a higher percentage of respondents experiencing light censorship encountered false information never or rarely (39.66\%) compared to heavily censored respondents (13\%). To investigate whether the impact of experienced censorship is associated with the frequency of encountering false information, we conducted a Chi-square test of independence. The test revealed a statistically significant association (\(\chi^2\) = 30.89, p-value = $3.2 \times 10^{-6}$), suggesting that censorship severity is significantly associated with the frequency of exposure to false information.

\noindent
{\bf Evading Censorship to Verify Information}.
We investigated whether the impact of experienced censorship is associated with the frequency of evading censorship to verify information by conducting a Chi-square test of independence between these two variables. The test revealed a statistically significant association (\(\chi^2\) = 38.69, p-value = $8.07 \times 10^{-8}$), suggesting significant differences in the frequency of evading censorship to verify information among participants who experienced light vs. heavy censorship. Figure~\ref{fig:evade} shows that indeed a higher percentage of lightly censored respondents never or rarely evaded censorship to verify information (70.49\%) compared to heavily censored respondents (27.08\%). Conversely, heavily censored respondents were more likely to often or always evade censorship (43.75\%) compared to lightly censored respondents (6.56\%).

In addition, we investigated whether the impact of experienced censorship is associated with the perceived difficulty of information verification by conducting a Chi-square test of independence between the perceived impact of worst censorship experienced and the difficulty of information verification. The results revealed a statistically significant association (\(\chi^2\) = 10.6, p-value = 0.031), suggesting significant differences in the perceived difficulty to verify information among participants who experienced light vs. heavy censorship. Figure~\ref{fig:verify} indeed shows a higher percentage of heavily censored respondents (82.52\%) who agree or strongly agree that censorship makes verifying information more difficult, compared to lightly censored respondents (76.27\%).

\section{Interview Findings}
\label{sec:interviews}

\begin{table*}[t]
\centering
\resizebox{\textwidth}{!}{ 
\begin{tabular}{lllllll} 
\toprule
ID & Gender & Age & Country of & Worst & Types of Censorship & Falsehood \\
& & & Censorship & Impact & Experienced & Frequency \\
\midrule
P1 & M & 26 - 35 & China & Extreme & Internet, Social media, Self censorship, Mass media & Sometimes \\
P2 & M & 26 - 35 & Ghana & Moderate & Internet, Self censorship, Mass media, Schools & Sometimes \\
P3 & M & 36 - 45 & Venezuela, USA & Extreme & Internet, Social media, Self censorship, Mass media & Always \\
P4 & F & 18 - 25 & USA& Slight & Self censorship, Schools, Research & Rarely \\
P5 & M & 26 - 35 & Iran & Extreme & Internet, Social media, Mass media & Often \\
P6 & M & 18 - 25 & Cuba, USA & Extreme & Internet, Social media, Self censorship, Mass media & Always \\
P7 & F & 26 - 35 & Iran, USA & Extreme & Internet, Social media, Self censorship, Mass media & Sometimes \\
P8 & M & 26 - 35 & China & Slight & Internet, Social media & Often \\
P9 & F & 18 - 25 & USA & Slight & Schools, Mass media & Sometimes \\
P10 & F & 18 - 25 & USA & Slight & Social media, Self censorship, Schools & Sometimes \\
P11 & M & 26 - 35 & India, SA & Moderate & Internet, Social media, Self censorship & Rarely \\
P12 & M & 26 - 35 & Iran & Significant & Internet, Social media, Self censorship, Mass media, Schools & Sometimes \\
P13 & M & 18 - 25 & China & Slight & Internet, Social media, Self censorship & Rarely \\
P14 & M & 18 - 25 & USA & Extreme & Internet, Social media, Self censorship & Sometimes \\
P15 & F & 26 - 35 & China & No impact & N/A & Sometimes \\
P16 & F & 26 - 35 & USA & Moderate & Social media, Self censorship, Schools & Sometimes \\
P17 & M & 26 - 35 & Iran, USA & Extreme & Internet, Social media, Self censorship, Mass media & Always \\
P18 & F & 18 - 25 & Venezuela & Moderate & Internet, Social media, Self censorship & Sometime \\
P19 & F & 26 - 35 & China & Slight & Internet, Social media, Mass media & Sometimes \\
P20 & M & 26 - 35 & Iran & Significant & Internet, Social media, Self censorship, Mass media, Research & Often \\
P21 & F & 36 - 45 & Iran, USA & Significant & Internet, Social media, Self censorship, Mass media & Often \\
P22 & M & 18 - 25 & China & Significant & Internet, Social media, Mass media & Rarely \\
P23 & F & 46 - 55 & China & Moderate  & Internet, Social media, Self censorship & Sometimes \\
P24 & F & 18 - 25 & Turkey & Extreme  & Internet, Social media, Self censorship, Mass media & Sometimes \\
P25 & F & 18 - 25 & Cuba & Extreme  & Internet, Self censorship, Schools, Mass media & Often \\
P26 & M & 18 - 25 & Colombia & Slight  & Social media, Self censorship, Mass media & Sometimes \\
P27 & M & 18 - 25 & Jordan, USA & Significant & Social media, Self censorship, Mass media & Rarely \\
P28 & M & 26 - 35 & Bangladesh & Significant & Internet, Social media, Self censorship, Mass media & Sometimes \\
P29 & M & 18 - 25 & USA & Extreme & Social media, Self censorship, Mass media & Always \\
P30 & M & 18 - 25 & Bangladesh & Extreme & Internet, Social media, Self censorship, Mass media & Often \\
\bottomrule
\end{tabular}}
\caption{Interview participant demographics, countries where they experienced censorship, survey-reported worst impact of censorship, types of censorship experienced, and frequency of encountering false information.}
\label{tab:interview:demographics}
\vspace{-15pt}
\end{table*}

Of the 121 survey respondents who expressed interest and provided contact information, 30 respondents who experienced censorship in diverse countries were invited to participate in the interview. All responded and consented. Table~\ref{tab:interview:demographics} shows participant demographics.

\subsection{Experiences with Censorship and False Information}

Our seven-month study (March–October 2024) covered key events in the participants' home countries. They include protests in Bangladesh~\cite{BDProtests} and Cuba~\cite{CUProtests} marked by internet shutdowns, and elections in the U.S.~\cite{USElectionInterference}, India~\cite{INDeepFakes}, Iran~\cite{IRBlocks}, Turkey~\cite{TurkeyElections}, and Venezuela~\cite{VZBlocks}, which saw heightened false information~\cite{USElectionInterference, INDeepFakes}, fact-checking site blocks~\cite{VZBlocks}, social media restrictions~\cite{TurkeyBlocks}, and media prosecutions~\cite{IRBlocks}.

Most participants reported experiencing access restrictions to various online content and services, including social media, news outlets, communication apps, and technical and educational resources. Participants from Bangladesh, Cuba, and Iran faced complete Internet shutdowns during political unrest, e.g., `{\it during the protests of 2021, the Internet was shut off for the entire island}'' (P25, CU). Several participants from Bangladesh were interviewed during the quota reform protests, when the authorities imposed a total shutdown of the Internet over multiple days. Other forms of restriction included bandwidth throttling~\cite{A13}, e.g., {\it social platforms were often slowed down to the extent that you couldn't load the page or access new posts}'' (P24, TR), and timed access, e.g., {\it they give you a specific half-hour slot per day when you can use the Internet}'' (P6, CU). The state is in a privileged position to implement these restrictions because ``{\it large-scale censorship requires significant effort and resources}'' (P5, IR). A few participants were more specific in naming censors, pointing out the Great Firewall of China and government ministries like the Information and Communication Technology (ICT) ministry in Bangladesh.

Participants also experienced enterprise-level censorship from social media platforms (both global and domestic replacements in China and Iran), mass media, employers, and education institutions, across both Global South countries and the U.S. Many participants also self-censor on social platforms, either to avoid government repercussions in Global South countries or to minimize digital footprints and avoid consequences from education institutions in the U.S.

Most participants encountered various types of false information, including misleading headlines, out-of-context reporting, propaganda, rumors and speculation, and conspiracy theories (e.g., on vaccines and climate change) in both Global South countries and the U.S. Participants from China, Iran, Cuba, and Venezuela were particularly exposed to state-driven disinformation, including misreported foreign crises and exaggerated government successes. Social media, influence operations, mass media, and educational institutions (e.g., history courses) were key sources of misinformation.

\subsection{Interplay of Censorship and False Information}
\label{sec:findings:interplay:censorship:io}

Many participants who experienced censorship in Global South countries like Iran, Venezuela, and Cuba, witnessed censorship being used in conjunction with the distribution of false information. For instance, participant P17, who experienced general Internet censorship in Iran, observed that ``{\it if there is false information, there is usually also censorship}''. This subsection investigates participant perceptions of the relationships between these information manipulation tools.

\noindent
{\bf Censorship Enables the Distribution of False Information}.
Participants who experienced significant impact of censorship in countries with state-level censorship, perceive that censorship enables false information. For instance, participant P3, who experienced censorship with extreme impact in Venezuela, observes that ``{\it the narrative controlled by censorship allowed the government to disseminate misinformation}''. For participant P5 who has experienced extreme censorship in Iran, this is due to the removal by censors of access to alternative sources of information, both external and domestic, ``{\it people may believe false information because they have no way to verify it against independent sources}''. This was corroborated by P6 who experienced extreme censorship in Cuba, ``{\it it's very hard to verify whether the news is true in Cuba, because there is only one source of news}''.

In addition to Internet restrictions, participants perceive that mass media outlets (TV stations, newspapers), provide common ground to censorship and false information. In particular, participants who experienced at least significant censorship in Bangladesh and China report that media outlets are a primary target of censorship. For instance, participant P28 who experienced censorship with significant impact in Bangladesh, observes that the government has the power to close down inconvenient news media outlets, e.g., ``{\it the government has shut down TV channels that were broadcasting information that the government did not want to be publicized}''. Participant P22, who experienced censorship of significant impact in China observed then that the remaining, accessible media outlets are often operated by the state, e.g., ``{\it the most accessible Chinese news stations are owned by the CCP}'' (P22, CN). Participant P6 also explained that the accessible media outlets implement censorship to comply with state requests, e.g.,

\blockquote{
``{\it [the media] only show you what they want you to see. They don't show the protests or the current revolution in Cuba, the widespread hunger}'' (P6, CU).}

Many of the interview participants who pointed to media outlets being major sources of false information, also reported frequently (always or often) encountering false information in the survey. They provided examples of false information distributed by such outlets, including Iranian media portraying the Black Friday event in the U.S. to be about people invading and robbing stores, and Chinese and Cuban media exaggerating reports about conflicts in the U.S. escalating into a civil war: ``{\it In Cuba they were saying there was a war going on in the U.S. I called my mom and sent her a video of what they were saying on the news}'' (P6, CU). The examples mentioned by participants build on real events but provide a distorted interpretation that aligns with state propaganda.

However, even in countries with more freedom of the press, several participants perceive that media outlets that distribute false information have financial links to certain entities that control the content they post. For instance, participant P11, who sometimes encountered false information in India, reports that the state provides news media with financial incentives to distribute government propaganda and false information, ``{\it the main [Indian] news channels spread misinformation because they are paid by the government}''. Participant P26, who sometimes encountered false information in Colombia, observes financial links of news media to political parties, ``{\it big families in Colombia that own news businesses are often related to political parties, making it difficult to trust the information they provide}''.

Participants who experienced false information in the U.S. perceive that U.S. news media outlets implement more subtle forms of censorship, through a combination of selective reporting of information and bias. In particular, when discussing the false information that he always encounters online in the U.S., participant P6 observes that in the U.S. it is not censorship, but the overwhelming amount of information, including false and biased information, that makes information verification difficult, ``{\it There's so much false information on the Internet that you don't know whether you would prefer to be censored or to have access to all that false information}''.

Censorship also targets academic research. Participant P20, who experienced censorship with significant impact in Iran, was not able to publish research results in national venues: ``{\it I was involved in research on [topic redacted to protect participant] in one province in Iran. We were not able to publish our scientific paper in Iranian journals because sharing such information is prohibited}'' (P20, IR).

Participant P20 further discussed efforts to organize a university congress and revealed how an organization in the Iranian censoring apparatus, i.e., the Security Council of Iran, has canceled it on the grounds that the congress topic might scare people.

In addition to censors that determine {\it what} research cannot be published, some censors determine {\it who} cannot access research results. In addition to participants who experienced Internet censorship that restricted their ability to access research and educational materials, participant P4 experienced restricted access to research publications by academic publishers. The participant, who experienced slight impact of censorship in the U.S., observes that publisher-imposed restrictions lead to situations where it is harder to fact-check online claims, ``{\it Paywalls or restricted access to academic studies can also act as a form of censorship, preventing people from accessing the information they need to verify what they've read or heard and distinguishing between credible sources and misinformation}'' (P4, US).

Several participants also observe that strict censorship may lead to speculation, where people make up stories or news based on the limited information available to them. Participant P24 who experienced censorship of extreme impact in Turkey further explained that censorship pushes people to seek information from unreliable sources,  ``{\it when people lose their access to legitimate information because it is censored, they are desperate to learn more about what is going on and look at other news sources, which are often not reliable}''. Participant P2, who experienced moderate censorship in Ghana, further revealed that this can lead to unreliable news sources exploiting censorship for their own benefit, and to the distribution of false information to a large audience,

\blockquote{
``{\it A news blogger claimed to have access to `The New York Times'. It turned out that he altered much of the content, deviating from the truth to draw more views}'' (P2, GH)}

Participant P2 explained that this blogger is a primary source of news of many people in Ghana, and is a news or media {\it smuggler}, who acquires censored information then sells it or distributes it in censored countries~\cite{NewsSmuggling}.

\noindent
{\bf False Information Enables Censorship}.
Participants who experienced censorship in diverse countries that include Iran, Cuba, Saudi Arabia, and the U.S., have witnessed censorship being used to prevent the distribution of false information. For instance, participant P11 who experienced censorship in Saudi Arabia, discussed censorship of pandemic-related misinformation, i.e., ``{\it during the COVID-19 pandemic, Saudi Arabia did a great job censoring disinformation and misinformation}''. This suggests that perceptions of the censor can affect interpretation of censorship. However, most participants were doubtful that censors remove only false information. For instance, participant P26 who often encountered false information online in Colombia, discussed how false information is used to justify censorship, ``{\it media controlled by political parties might claim to block `fake news', but in reality they could be silencing valid criticism}''. This is echoed by participants who experienced censorship on social media. For instance, participant P24 who experienced extreme impact censorship in Turkey, discussed social platforms removing true but inconvenient content and even suspending accounts that interact with the content:

\blockquote{
``{\it I was banned from Instagram for a month because I interacted with posts with hashtags related to the wildfires}'' (P24, TR)}

Several participants, including P6, who experienced extreme impact censorship in Cuba, even perceive that global social platforms collaborate with state censors to remove content that does not necessarily violate their terms of service, ``{\it Facebook and Instagram work in Cuba but you won't see as much content as you would in the U.S. And if you tried to post videos from Cuba, they would be taken down immediately}'' (P6, CU).

\subsection{Interplay of Influence Operations and Censorship}

\noindent
{\bf Influence Operations and Censorship on Social Media}.
A few participants witnessed operations conducted by foreign organizations, in particular Russian operations conducted by trolls and bots that distribute false information, promote conspiracy theories, e.g., about vaccines and climate change, and generally seek to ``{\it generate confusion and fear among the population, destabilize the nation, and undermine trust in the government and institutions}'' (P3, VE). Most participants however discussed domestic operations. For instance, participant P22 who experienced significant censorship in China attributes influence operations to the state, ``{\it the government uses various groups and state-affiliated organizations to spread false narratives to maintain control and influence public opinion}'' (P22, CN). Some participants named the responsible organizations, e.g., Taiwan’s 1450 Internet army and Iran's Basij~\cite{Basij}, and revealed detailed knowledge of their operations.

Participants who experienced heavy censorship in China and Iran report monitoring and censorship implemented by domestic social media platforms that are coerced to collaborate with state censors. Participant P1, who experienced extreme impact censorship in China further observes that social media companies in China often impose stricter rules than the government, to avoid penalties levied on non-compliant platforms. Chinese platforms implement censorship through keyword filtering, removing posts containing sensitive words~\cite{KR21, XK19, VP15}. Participant P1 further explains that users have to carefully consider their posts since even unintended combinations of words that could be perceived to be sensitive, might trigger censorship. On global social media, participants experience censorship through post removal, shadow banning, and account suspensions, ``{\it In Instagram, I posted about a political issue that occurred a few months ago, and others shared a post that the platform decided shouldn't be spread, even if it's true. [Instagram] will delete the story, post, or any content they prefer not to have shared}'' (P7, US). Participants complain that content moderation does not provide them feedback and recommend that platforms provide more transparent interventions to address perceptions of censorship. 

\noindent
{\bf Cooperation Between Censorship and Influence Operations}.
Several participants report cooperation between influence operations and censorship efforts in their countries. For instance, participant P20 who experienced significant impact censorship in Iran observes that 

\blockquote{
``{\it In Iran, influence operations always go hand in hand with censorship. The government wants to spread its information, and influence operations are used to amplify this message while censorship ensures that opposing views are suppressed. They are always aligned with each other}'' (P20, IR)}

Participants perceive that most of the information distributed by influence operations is false. They gave examples that include misinformation about the impact of Iran's attacks on Israel, that was associated with pictures and videos, and used hashtags and amplification to give the illusion of popularity. Participants from Iran who witnessed influence operations on social media observe that they are used not only to distribute false information and propaganda, but also to divert attention, ``{\it they present unimportant news to get the focus away from the main events}'' (P5, IR), disrupt discussions and suppress dissenting opinions ``{\it if someone is criticizing the government, they intervene to disrupt the conversation}'' (P12, IR). Participant P29 who experienced censorship of extreme impact in the U.S. and always encounters false information online, notes that ``{\it when influence operations drown out the real news, they indirectly censor it}''.

Further, and similar to false information, several participants, including some who experienced censorship in China, observe that influence operations can be used to justify censorship. In particular, P8 who experienced slight impact censorship in China and often encounters false information online, points out that ``{\it the significance of preventing access to foreign operations and news outweighs censorship risks}'' (P8, CN). Influence operations also work together with censorship to create information cocoons, by posting similar information from seemingly different sources, ``{\it when a person sees the same false news on 3-4 different resources, they are more likely to believe it as true without checking for outsider sources}'' (P5, IR). This exploits vulnerabilities in information verification strategies that check the consistency of reports, described in $\S$~\ref{sec:findings:interplay:evasion}.

\noindent
{\bf Antagonism of Influence Operations and Censorship}.
Some participants who witnessed influence operations in China and Iran also presume that some operations critical of the government are conducted by foreign organizations. For instance, participant P20 from Iran notes that ``{\it I see systematic comments on specific subjects and I assume they are from influence operations opposing the government that are based in other countries}'' (P20, IR). This reveals perceptions that influence operations are able to evade censorship.

This is echoed by a few participants who experienced censorship in Latin American countries, who observe that influence operations can be used to combat censorship. For instance, P18 who experienced moderate impact censorship in Venezuela suggests that influence operations can ``{\it use covert methods to bypass restrictions and reach a wider audience, spreading their narratives despite attempts to censor them}''. Participant P26 who experienced slight impact censorship in Colombia, proposes that censorship evasion solutions leverage techniques used by influence operations to involve regular users, ``{\it information and news was spread by asking you to share the message with ten people to avoid being blocked}'', perhaps in an attempt to initiate grassroots campaigns.

\subsection{Information Cocoons: Creation and Evasion}
\label{sec:findings:interplay:evasion}

For a few participants from Cuba and China, the combination of censorship and false information was so effective that it created cocoons of false information and censorship. For instance, participant P6 who grew up experiencing extreme censorship in Cuba, believed the government propaganda:

\blockquote{
``{\it In the beginning it was hard to tell because I was born and raised in Cuba, where the news always blamed the U.S. for everything wrong in Cuba. I believed it at the time. But when I left Cuba I started seeing things from an outside perspective. I realized everything Cuba was saying about the government, how they treat people, and about the U.S., was a complete lie. Over time, I understood that everything I was taught from childhood to adulthood in Cuba was a lie}'' (P6, CU)}.

Trust in government propaganda led to participants not being aware of these cocoons, which therefore they did not seek to evade. Participant P22 who experienced significant censorship in China even explains that escaping the cocoons may only be possible when physically removed from them. Participant P15 who experienced no impact of censorship in China, reveals that her experience of information cocoons was enabled by her exclusive use of Chinese social media.

\noindent
{\bf Evading Censorship}.
Most interview participants have evaded censorship, using either free or paid VPNs. Several participants from Iran and China discussed the use of official VPNs that are provided and tolerated by the censors, e.g., ``{\it the government provides VPNs specifically designed for use within [Iran] that allow access to social media but block access to more sensitive sites}'' (P7, IR). Participant P7 experienced censorship of extreme impact in Iran. This relates to findings from other participants who reported that censors monitor user activities on social media, even after they evade censorship and are outside the country.

However, a few participants further evade such monitoring, by hiding meaning in seemingly innocuous communications, e.g., ``{\it Rather than writing directly about sensitive topics I had to convey my thoughts in a coded or metaphorical manner, through poetry and subtle movie themes}'' (P7, IR). Participant P1, who experienced censorship with extreme impact in China, also evaded keyword filtering on domestic platforms by using jargon and homonyms, ``{\it the necessity to navigate around keyword filtering systems has led to the development of inventive ways to discuss topics without directly triggering censorship mechanisms}'' (P1, CN, US).

Participants who experienced Internet shutdowns find it almost impossible to evade censorship. However, participant P28 reported that in Bangladesh this is possible for some, ``{\it People living near the border with India had some alternatives. Some of them had Indian SIM cards and could connect to Indian networks if they went close to the border}'' (P28, BD). This strategy is similar to the one employed by participants who experienced lighter forms of censorship implemented on employer or school devices in the U.S., e.g., ``{\it I use my cell phone to connect to the sites and social platforms censored by my employer}'' (P3, VE, US).

Surprisingly, most participants who experienced significant or extreme censorship in the U.S. did not attempt to evade it, but instead choose to practice self-censorship. While users can create pseudonymous accounts or use safer alternatives like ``{\it Telegram because I've never seen anything get taken down there}'' (P14, US), participants prefer to remain on mainstream platforms where their social contacts are active. Other reasons include perceptions of censorship being more subtle in global platforms, coupled with more trust in their moderation which users tend to interpret less like a personal threat.

\noindent
{\bf Detecting and Avoiding False Information}.
To evade false information, many participants use reputable information sources and avoid sources known to distribute false information, including state-controlled outlets. Several participants also verify the consistency of information from multiple sources, ``{\it I try to use media that is less directly related to the government. If several reputable sources say the same thing, I start to believe it}'' (P5, IR). However, participant P9 who sometimes encounters false information in the U.S., reveals a vulnerability that occurs when all (or multiple) available sources present the same but incomplete or incorrect information, ``{\it all these different news sites were saying the same thing, making it difficult to find the actual truth}''.

Several participants have evaded censorship to access alternative sources of information that allow them to verify information promoted within the censored region. In addition to using VPNs in order to access outside news media, a few participants reach out to people located close to the reported events to seek peer validation. For instance, participant P6 explained that in Cuba it is almost impossible to verify information unless you know someone outside the country, and gave an example:

\blockquote{
``{\it When I was in Cuba, they were saying there was a war going on in the U.S. I got really worried because I had my family over there. I called my mom and asked, `What's happening? Is there really a war going on?' and sent her a video of what they were saying on the news. She reassured me that it wasn't true at all and that nothing was happening.}'' (P6, CU)}

This refers to a border standoff between Texas and the federal government that escalated in early 2024, and was used by the Cuban media to promote the news of a civil war in the U.S. Participant P21 who experienced censorship in Iran of significant impact also contacts people she trusts that personally know the source of the information, ``{\it If it's about a government meeting, I ask people and check with my reliable sources. These are people I know who keep in touch with the news and know the journalists who write the articles. They are very detailed, so I can double-check with them}'' (P21, IR).

To further avoid monitoring and censorship over online platforms like social media or email, P21 uses offline channels, e.g., landline phones, to communicate with such personal contacts. This includes cases where the news concerns events alleged to have taken place in other parts of Iran, when P21 communicates with personal contacts located in those areas. Participant P2 who experienced moderate censorship and sometimes encountered false information in Ghana, relied on news smugglers not only to access censored information but also to verify information:

\blockquote{
``{\it I verified the authenticity of the news after a few days when someone else paid for access and shared the information}'' (P2, GH).}

Such sources of information are, however, not official. Using personal contacts to obtain and verify information has drawbacks, and can help the spread of false information, as mentioned earlier by participant P2 when he discussed how news smugglers also create false information designed to attract interest and views.

\section{Discussion}
\label{sec:discussion}

\begin{figure}
\centering
\includegraphics[width=0.63\columnwidth]{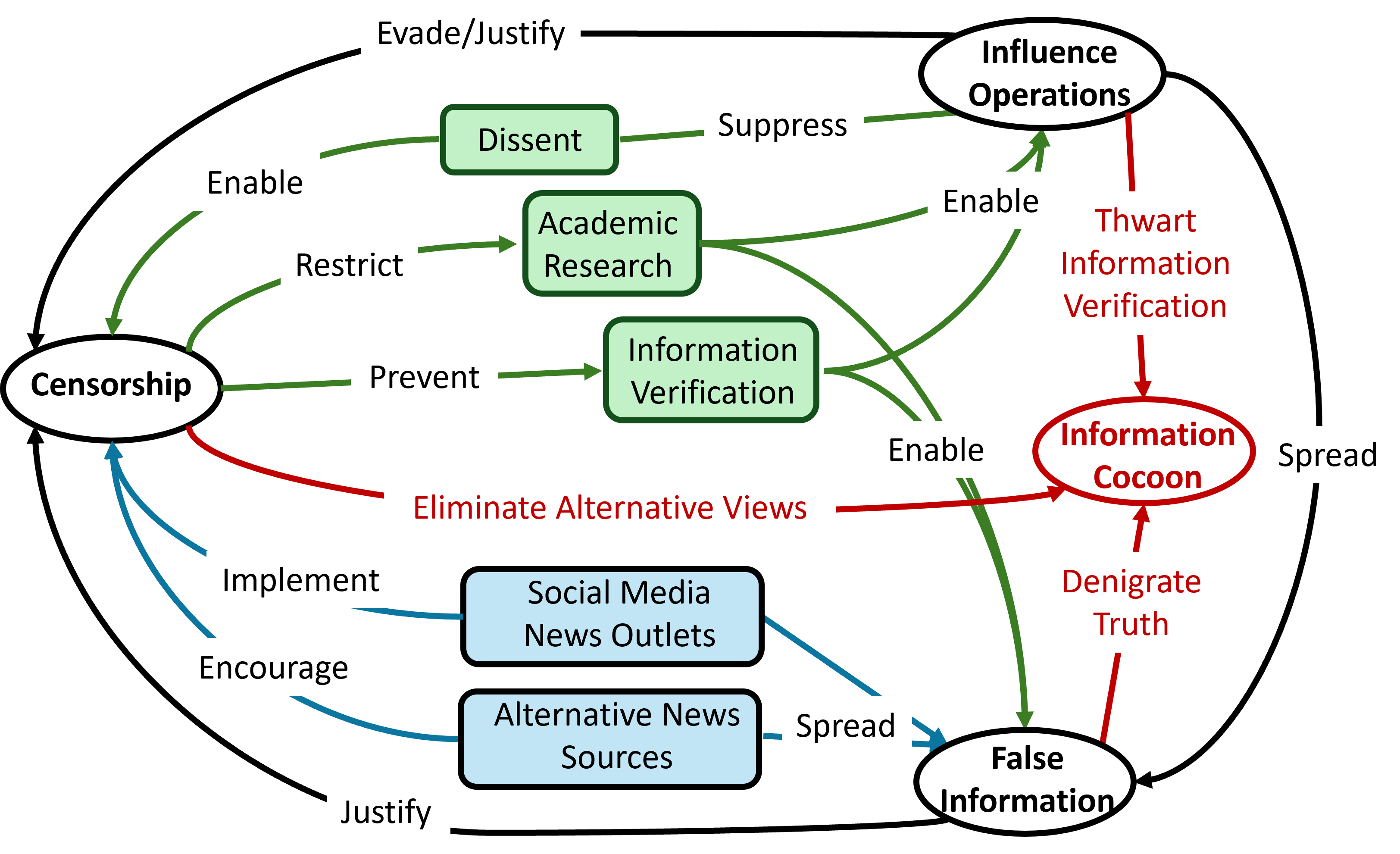}
\caption{Map of relations between censorship, false information and influence operations, and their use in creating information cocoons.}
\vspace{-15pt}
\label{fig:censorio:map}
\end{figure}

\subsection{Censorship, False Information, Influence Operations: Interplay and Evasion}
\label{sec:discussion:interplay}

Figure~\ref{fig:censorio:map} illustrates key study findings on the interplay of information manipulation tools discussed next.

\noindent
{\bf Censorship Facilitates False Information and Influence Operations}.
Our findings support Mou et al.’s~\cite{MAF13} conclusion that censorship in China can foster more critical and intentional news consumption, and show this effect extends beyond China. We further find that when users believe censors actively promote false information, they are motivated to evade censorship to validate claims. This aligns with survey results showing that users aware of extensive censorship more frequently engage in evasion to verify information.

Among survey respondents who experienced censorship, 79.6\% agreed that it hampers verification, rising to 85.1\% for those under heavy censorship. By obstructing fact-checking and restricting academic research and access to it, censorship enables false information and influence operations (green arrows in Figure~\ref{fig:censorio:map}). For instance, paywalls by academic publishers~\cite{H20, ElsevierProfit, SpringerNatureProfits} hinder access to credible sources, particularly for users in developing countries, despite institutional subsidies in wealthier contexts. Consequently, censorship that restricts access to scholarly work is perceived to worsen misinformation by preventing users from verifying facts.

Censorship also fuels false information by prompting speculation and reliance on unreliable sources (blue arrow from censorship to the false information node). This dynamic suggests a vicious cycle in which censorship supports false information, which in turn leads to further censorship. Conversely, the concept of false information is used by censors to justify their actions, e.g., by coercing social platforms to remove flagged content (black arrow from the false information to the censorship node).

\noindent
{\bf Relationship of Influence Operations to Censorship}.
Influence operations often spread false information and are perceived to facilitate censorship by diverting attention from important events, disrupting discourse, and suppressing dissent (green arrow from the IO to the censorship node in Figure~\ref{fig:censorio:map}). These tactics, documented in China~\cite{KPR17}, align with censorship practices in their shared objective to obstruct authentic social media communications.

Influence operations are also perceived to be a means to evade censorship, with participants suggesting flash grassroots campaigns that rapidly circulate sensitive content before censors can respond (black arrow from the IO to the censorship node). However, similar to false information, censors also leverage influence operations to justify actions, e.g., to counter foreign operations. This dynamic suggests public approval of censorship may influence its perceived impact, explaining why some survey respondents reported minimal censorship in countries like China, see Figure~\ref{fig:country:impact}.

\noindent
{\bf Information Cocoons: Cooperation Between Information Manipulators}.
Prior work on {\it information cocoons}, including Sunstein’s {\it daily me}~\cite{S01} and Pariser’s {\it filter bubble}~\cite{P11}, frames them to emerge from user self-selection and algorithmic personalization in open media environments. These frameworks have shaped CSCW discussions of echo chambers and ideological clustering online~\cite{CDGQS21, TGBVSSSN18}. Our findings extend these concepts by showing how information cocoons (red node in Figure~\ref{fig:censorio:map}) can be deliberately constructed through coordinated information control in settings shaped by state censorship, influence operations, and infrastructural manipulation. This reconceptualization shifts the analytic focus from accidental exposure to intentional epistemic barriers, which Roberts calls {\it infrastructural violence}~\cite{RO12}.

Participants from countries including Cuba, China, and Iran described environments where censorship was not merely suppressive, but constitutive, actively shaping the information landscape by limiting dissenting views and enabling the dissemination of targeted falsehoods.
%
%
Such information control systems created cocoons of false narratives that were difficult to detect from within, especially when trust in the information infrastructure remained intact. Several participants only recognized the extent of manipulation after leaving these environments, revealing how physical and informational isolation reinforced belief in state propaganda.

Censorship restricts alternative perspectives and controls narratives (red arrow from censorship to information cocoon in Figure\ref{fig:censorio:map}). Participants also noted that influence operations exploit common verification strategies like cross-source consistency, to reinforce those narratives by spreading identical or near-identical false content~\cite{VNCCGL24} from seemingly diverse sources, manufacturing the illusion of consensus. Other tactics include disrupting discourse by overwhelming platforms with noise, diverting attention from dissent, and jamming communication channels, consistent with information warfare strategies~\cite{ODIN}.

Our study expands CSCW understandings of information cocoons by emphasizing their systemic, coordinated construction in authoritarian settings, particularly in the Global South, where participants encountered tight platform control, state-enterprise cooperation, and limited access to trustworthy information. We also document how users resist these systems using informal verification, metaphorical language, trusted social networks, and cross-border connectivity. These practices reveal that user agency is not erased but reconfigured, shaped by relational trust, creative adaptation, and unequal access to resources. Our findings offer a conceptual extension of information cocoons by situating them within a broader socio-technical system of control, that integrates censorship, false information, and influence operations. This introduces a global and infrastructural perspective to debates on information integrity, moderation, and participation, and complements prior analyses of censorship tactics~\cite{KPR13, KPR17} by emphasizing user agency and resistance within controlled environments.

\noindent
{\bf Case Studies}.
The cooperation between organizations spreading false information through influence operations and those coercing censorship on social platforms and news media can be so effective that some participants experienced only mild censorship in countries with low Internet freedom~\cite{FREEDOMNET} and realized they were living in information cocoons only after physically removed from them. For example, China, with an Internet freedom score of 9/100 in 2024~\cite{ChinaFreedom}, has created information cocoons through an advanced censorship infrastructure, state-led propaganda, and domestic alternatives to social media. The widespread availability of domestic content on these platforms reduces the need for users to access blocked international platforms, especially since these platforms are rarely used by their social contacts. Similarly, Cuba, an island nation with an Internet freedom score of 20/100 in 2024~\cite{CubaFreedom}, creates an invisible cocoon through a combination of physical isolation~\cite{CubaIsolation} and digital isolation reinforced by strong state-controlled propaganda~\cite{DAB16}.

In contrast to China, Iran lacks the same level of technological oversight and faces challenges in controlling information due to widespread censorship circumvention efforts~\cite{ANYB24}, and more active underground media and opposition~\cite{R17}. In contrast to Cuba, efforts in Bangladesh to impose information cocoons through total Internet shutdowns~\cite{BDProtests} were only temporarily effective, due to strong student movements~\cite{J21} and proximity to countries with greater Internet freedom~\cite{IndiaFreedom} that prevented the invisibility of censorship and false information.

\subsection{Designing for Censored Environments}
\label{sec:discussion:design}

\noindent
{\bf Resilience through Technology: Escaping Information Cocoons}.
Censorship can be overt, e.g., embedded in platform design or connectivity constraints, or internalized through social norms and fear~\cite{RO12, YL25, GS23}. Our findings extend prior knowledge of active resistance to censorship, by detailing specific adaptive strategies employed in heavily censored contexts like Iran, China, and Bangladesh. Participants described using VPNs, coded language to evade keyword filtering, and physically accessing foreign networks. These tactics not only circumvent restrictions but also reveal awareness of information control mechanisms. Building on Yao and Li~\cite{YL25}'s notion of micro-foundational mechanisms that support authoritarian resilience (e.g., self-censorship, avoidance), we identify parallel {\it micro-level adaptive strategies}, e.g., metaphor and homonym use, relying on strong-tie networks, that reveal further creative resistance to censorship. This contributes to CSCW by showing how users negotiate and reshape repressive socio-technical systems.

Efforts to evade information cocoons align closely with the CSCW concept of {\it resilience}~\cite{MS08, MAS09, VD17, S19, PKVZ23},  the capacity to adapt and maintain functionality despite disruptions and environmental constraints~\cite{KW03}. Extending prior work on information technology use during disruption~\cite{MS08, MAS09}, we show people create new routines to evade censorship and verify information. Resilience manifests individually, e.g., embedding meaning in innocuous messages, and collectively, through shared strategies like VPN use, coded language, and relying on ``news smugglers''. This reveals community-driven verification dependent on strong social ties~\cite{VD17}.

We now distill design implications for platforms in these contexts.

\noindent
{\bf Design Recommendations for Content Moderation and Verification}.
Platforms moderate content to comply with censorship requests~\cite{XTurkey, TurkeyDisinformation, TurkeyCensorship, L21, RelyFactCheck, XIndiaBlock}, and with policies~\cite{FacebookMisinformation, TwitterMisinformation} and regulations~\cite{EUDSA, GDPRChildren, COPPA}. However, moderation can be perceived to be censorship.
This leads to {\it design recommendation 1}: platforms should increase transparency in moderation practices to reduce perceptions of censorship.

Participants verified information through social contacts near the source, e.g., relatives abroad verifying news for those in censored countries. This motivates {\it design recommendation 2}: platforms should facilitate claim verification via social contacts with more accurate information. Further, participants used multiple sources to verify claims. This leads to {\it design recommendation 3}: platforms should promote awareness of and access to diverse viewpoints on posted content.

\noindent
{\bf Designing for Privacy and Plausible Deniability in Censored Environments}.
Participants from Iran used steganography to conceal meaning and feared surveillance despite circumvention efforts. Participants also discussed cases where authorities accessed social media accounts of detainees by confiscating their devices and using (1) their posts to incriminate them, and (2) their networks to target social contacts~\cite{IranTelegram}. This motivates {\it design recommendation 4}: platforms must support privacy and plausible deniability, enabling users under coercion to reveal benign explanations for their social media activities while evading censorship.

\subsection{Appropriation and Counter-Appropriation}
\label{sec:discussion:appropriation}

We draw on the concepts of appropriation and counter-appropriation~\cite{WRAR22, LKMRW19, SP18} to examine the co-evolution of strategies developed by censors and distributors of false information, and the responses developed by users. Appropriation, i.e., the adoption of new practices around emerging technologies~\cite{SP18}, can be used to describe evolving tactics by censors and influence operations. Counter-appropriation describes user adaptations in response to these constraints. Our findings extend CSCW frameworks by showing how these dynamics unfold in diverse socio-political contexts, and how platform affordances shape both sides of this arms race.

\noindent
{\bf Censorship}.
Domestic platforms in China and Iran illustrate appropriation by censors to reinforce information cocoons and suppress dissent. Replacing banned international platforms with state-controlled alternatives enables easier monitoring, takedown enforcement, and message shaping, while obscuring censorship strategies from outside observers. Participants reported this to be a direct response to earlier counter-appropriation efforts like VPN use and foreign platform access.

State-approved VPNs discussed by participants reflect a further stage of appropriation, where censorship infrastructure leverages evasion tools to regain surveillance capability~\cite{VPNOwnership, SensorTower, IVSKP16, KDVSKV18}. These developments extend prior critiques of tool-based resistance~\cite{SHL14, BSBD15, D14} by illustrating how adversaries now appropriate resistance tools themselves, transforming them into surveillance vectors, and by revealing how users adapt when even trusted tools become compromised. In turn, participants described evolving counter-appropriation strategies, e.g., reverse-engineering filters and using covert channels on tolerated platforms. This reveals how resistance persists even under tightened constraints, and suggests the need for platform designs that support context-specific user tactics, instead of assuming universally safe tools.

\noindent
{\bf False Information and Influence Operations}.
Participants across countries engaged in counter-appropriation to resist misinformation, using out-of-band communication, source vetting, and cross-verification. However, these strategies have triggered appropriation responses from influence operations, including large-scale content duplication~\cite{VNCCGL24}, where identical posts appear across different accounts to simulate consensus. This subverts user verification strategies by exploiting perceived consensus.

These findings contribute to CSCW by revealing a recursive interplay between how users access, interpret and evaluate information and adversarial appropriation of platform affordances. They also reveal the influence played by socio-political environments on resistance strategies, and the importance of context-specific interventions.

\subsection{Comparison of Perceptions and Experiences in the U.S. vs. Global South}
\label{sec:discussion:comparison}

Our study reveals both similarities and differences in how censorship and false information are perceived in the Global South and the U.S. Many U.S.-based participants reported experiencing censorship domestically, while participants from the Global South often viewed censorship in the U.S. to be less severe than in their home countries. However, some participants also experienced high-impact censorship in the U.S., while a few participants from repressive regimes were unaffected.

Across countries, social media and news outlets are responsible for information manipulation albeit through different strategies. In many Global South countries, censorship hinders information verification. In contrast, U.S. participants struggled to verify content due to the abundance of biased and false information from multiple, seemingly consistent sources. This undermined participant-reported content verification strategies like cross-checking reports. In addition, while academic research is censored in some Global South countries, access to scholarly results is often restricted globally by Global North publishers~\cite{H20, ElsevierProfit, SpringerNatureProfits, JSTORHack}.

U.S. participants reported censorship on social media through post removal, shadow banning, and account suspensions, typically in response to flagged misinformation~\cite{FacebookMisinformation, TwitterMisinformation} or regulatory compliance~\cite{EUDSA, GDPRChildren, COPPA}. In some Global South countries, platforms remove content flagged by government-appointed fact-checkers, or in response to court and state demands~\cite{XTurkey, TurkeyDisinformation, TurkeyCensorship, L21, RelyFactCheck, XIndiaBlock}.

In addition, our study extends prior work focused on China by providing a broader, cross-national perspective on user experiences and responses to information manipulation. Extending Lu et al.~\cite{LJLNW20} on astroturfing and comment manipulation, we observe similar dynamics in countries like Iran, Cuba, and Venezuela, where censorship and misinformation jointly shape narratives and suppress dissent (e.g., P3, VE, and P6, CU). We also find parallels to the epistemic and emotional harms documented in China by Li and Nardi~\cite{LN21}, where participants in Bangladesh and Iran describe Internet shutdowns, blocked academic publications (P20, IR), and psychological distress under state-controlled media.

Importantly, our findings show that these harms and manipulation strategies are not unique to China but appear in various forms across the Global South and even the U.S. participants described more subtle censorship, e.g., algorithmic suppression on social media and workplace-imposed restrictions, leading to self-censorship or topic avoidance (e.g., P3, P14). We also identify how systems of censorship, false information, and influence operations co-evolve and reinforce each other across diverse national contexts. This allows us to identify not only mechanisms of manipulation but also adaptive strategies. Across countries, participants revealed diverse strategies that include VPN use and coded communications in China and Iran, and verification through personal networks in Cuba, Ghana and Iran. These findings reveal the role of user agency in navigating complex information environments and suggest the need for transparency interventions sensitive to and grounded on local context.

\section{Design Implications}
\label{sec:design:implications}

We use the design recommendations from Section~\ref{sec:discussion:design} to recommend techniques for social platforms to help users navigate information cocoons in two different operational settings.

\subsection{Interventions to Navigate Information Manipulation}
\label{sec:design:interventions}

We focus first on settings where global platforms like Meta and X are accessible in countries governed by different laws requiring content removal. In Section~\ref{sec:discussion:comparison} we discuss findings that users face various barriers to trustworthy information, including tightly controlled media ecosystems in several Global South countries and algorithmic suppression and information overload in the U.S. These differences lead to adaptive, context-specific user strategies. This suggests transparency interventions should avoid one-size-fits-all solutions and support users’ context-sensitive practices by explaining moderation decisions and offering trustworthy alternatives. This motivates design interventions that provide transparency adaptively, instead of relying solely on content removal or account suspensions.

Building on design recommendation 1  (Section~\ref{sec:discussion:design}, increasing moderation transparency) and our discussion of the interplay between false information and censorship (Section~\ref{sec:discussion:interplay}), we propose new interventions to help users distinguish between content removed due to policy violations or censorship and content flagged for misinformation or other harms (e.g., hate speech). We use the term {\it sensitive} to refer to such content.

Instead of removing sensitive content, platforms could extend the warning-based approaches (e.g., by Meta and X~\cite{FacebookWarning, TwitterWarning}) that blur or cover content with context warnings. These warnings should clarify the basis for intervention, e.g., policy violations, promotion by influence operations, untrustworthy sources or references, or takedown requests. To enhance transparency, warnings should link to supporting evidence like fact-checks, policies, applicable laws, or details on influence operations and sites promoting the content. To further enhance transparency, users should also be informed of potential consequences for interacting with such content, e.g., account penalties.

In line with design recommendation 2 (facilitating claim verification), we propose nudging users to consult social contacts with relevant context before engaging with sensitive content. Platforms could support this by inferring contact location from profile data. Consistent with design recommendation 3 (exposing diverse viewpoints), we suggest emphasizing disagreements among sources or contacts about a post's accuracy and presenting alternative views.

Finally, we propose an intervention to warn users when they view posts from accounts linked to influence operations or known for spreading falsehoods. If the user has a weak tie to the account, the platform can prompt disengagement through actions like muting, restricting, or unfollowing~\cite{InstagramActions, XMute}. If the tie is strong, e.g., a friend or family member, the platform can instead encourage users to privately challenge the content’s accuracy.

These interventions seek to promote transparency and context, and raise user awareness of the interplay between false information, influence operations, and censorship. Instead of removing content or suspending accounts, these interventions shift the responsibility from platforms by empowering users to decide whether to engage, thus reducing the influence of false or manipulated information through informed choice.

\subsection{Plausibly Deniable Censorship Evasion Through Social Platforms}
\label{sec:design:plausible}

We focus now on a setting where global platforms like Meta and X are accessed by users in countries like Iran and China (after evading censorship using tools like VPNs~\cite{NordVPN, ExpressVPN} and Tor~\cite{Tor}), and, more recently, even the U.S.~\cite{PhoneInspections, USNPhone}. Building on design recommendation 4 (Section~\ref{sec:discussion:design}), we propose that platforms help users maintain privacy and plausible deniability of social media activities, even when coerced by authorities to grant access to devices and accounts.

Extending Meta's feature for multiple profiles under one account, we respond to participant accounts of evading censorship by maintaining multiple identities, switching platforms, or hiding sensitive communication (e.g., P1, CN; P7, IR; P14, US). These strategies were often described to be time-consuming, risky, and dependent on informal tools or practices. To address these limitations, we propose platform-supported {\it dual-profile accounts}, allowing users to create and switch between distinct public and private profiles within a single account.

Each profile has the same number of social contacts, communications, and data items. The private profile contains the user’s real contacts and content, while the public profile displays dummy versions. Logging in with the public password shows only the public profile, while the private password grants access to both. When a user adds a sensitive contact in the private profile, the platform uses generative AI~\cite{SocialAI, OpenArt, ChatGPT} to mirror this action in the public profile by generating a benign, dummy contact.

Similarly, when sensitive content is posted in the private profile, AI tools generate same-size, non-sensitive versions for the public profile. LLMs can detect when the original content would be censored and produce a safer alternative. This mirrors participant strategies of using coded content to avoid censorship (e.g., P1, CN) and reduces the burden of managing dual narratives manually.

This design allows users to evade censorship while maintaining a plausible, innocuous account activity log. If coerced, they can reveal the public password and plausibly deny the private one. Even if authorities monitor network traffic, the mirrored activities ensure alignment between observed behavior and public profile content.

\subsection{Limitations}

The proposed interventions and techniques remain speculative without implementation and evaluation. Understanding user perceptions requires studying individuals regularly exposed to false information and censorship. Effective assessment should consider real-world complexity, adaptability across users and technologies, and ethical implications.

\noindent
{\bf Ethical Considerations}.
The proposed solutions raise ethical concerns for both platforms and users. Allowing users to censor or evade censorship risks detection by authorities, both in repressive regimes and democratic countries~\cite{AppleEncryption, NSASpy, FacebookDisclosure}. Platforms may face financial penalties, bans that impact all users, or legal obligations to disclose data. This could endanger vulnerable users. In addition, access via anonymizers enables anonymous accounts, which can be exploited to spread harmful content.

\section{Conclusions}

In this paper we explored perceptions and experiences of participants across diverse countries regarding how censorship, false information, and influence operations interact and shape each other. Our study reveals that the type and impact of censorship influence the frequency of encountering false information, and efforts and challenges to evade censorship to verify information. Our findings highlight experience-based perceptions that censorship facilitates false information, both by censors and third-party opportunists, while influence operations propagate false information and support censorship efforts. We leverage study insights to propose interventions that raise user awareness to the interplay of influence manipulation tools and platform-assisted techniques to plausibly deniably evade censorship.

\section{Acknowledgments}

This work was supported in part by the National Science Foundation under grants 2321649, 2114911, 2013671. We also thank the anonymous reviewers and coordinators for their helpful feedback.

\bibliographystyle{ACM-Reference-Format}
\bibliography{censorship,crackdown,factcheck,misinformation}

\appendix

\section{Survey}
\label{appendix:survey}

\subsection{Introduction}

You are invited to participate in a research survey on internet censorship and false information. The purpose of this survey is to understand your experiences with internet censorship in specific countries and your exposure to online false information. Your participation is entirely voluntary.\\

\noindent
In the following, Internet censorship refers to the control or suppression of online information, which may involve blocking access to certain websites, filtering or removing content, and restricting the ability of individuals to communicate freely online. Censorship can involve a variety of sites and services that include:

\begin{compactitem}

\item
Social Networks (for example, Facebook, Twitter, TikTok)

\item
Communication Apps (for example, WhatsApp, Telegram, Signal, WeChat, Line)

\item
Search Engines (for example, Google Search, DuckDuckGo)

\item
Email Services (for example, Gmail, Outlook, ProtonMail)

\item
News Websites (for example, BBC, CNN, Al Jazeera)

\item
Cloud Storage Services (for example, Google Drive, Dropbox, iCloud)

\item
Cryptocurrencies (for example, Bitcoin, Ethereum)

\item
Online Payment Services (for example, PayPal, Stripe, Square, Wise)

\item
VPN Providers (for example, NordVPN, ExpressVPN)

\end{compactitem}

In addition, false information refers to content that is inaccurate or misleading, whether spread intentionally (for example, propaganda, fabricated or manipulated content) or unintentionally (for example, factual mistakes, unverified claims).

\noindent
\textbf{NUMBER OF STUDY PARTICIPANTS}  

\noindent
If you decide to participate in this study, you will be one of up to 1000 people.  

\noindent
\textbf{DURATION OF THE STUDY}  

\noindent
Your participation will require between 2 – 5 minutes.  

\noindent
\textbf{DATA COLLECTED}  

\noindent
In this study, we will collect your responses to the survey questions. We will store responses indexed under pseudonymous identifiers.  

\noindent
\textbf{CONFIDENTIALITY}  

\noindent
Your study records will be kept confidential, and protected to the fullest extent allowed by law. Any published reports will not contain information enabling subject identification. Your identity is safe as we do not collect personally identifiable information.  

\noindent
\textbf{PARTICIPANT AGREEMENT}  

\noindent
I have read the information in this consent form and agree to participate in this study.

(Agree/Disagree)

\subsection{Questions}

1. I have experienced Internet censorship while living in the following countries. If the answer is ``I did not experience censorship in any country'', ask 1.1 and 1.2 then stop the interview.

1.1. I have encountered information online that I suspected was false. 

\newmaterial{(Never, Rarely, Sometimes, Often, Always)}

1.2. Censorship makes it harder to verify false information that circulates online. 

\newmaterial{(Strongly Agree, Agree, Neither agree nor Disagree, Disagree, Strongly disagree)}

\noindent
2. While living in a country with Internet censorship I have encountered information online that I suspected was false.

\newmaterial{(Never, Rarely, Sometimes, Often, Always)}

\noindent
3. I have evaded censorship in order to verify information I found online.

\newmaterial{(Never, Rarely, Sometimes, Often, Always)}

\noindent
4. Censorship makes it harder to verify false information that circulates online.

\newmaterial{(Strongly Agree, Agree, Neither agree nor Disagree, Disagree, Strongly disagree)}

\noindent
5. Outside of the US, the worst Internet censorship I experienced in a country with censorship had
(1) No impact at all
(2) Slight impact
(3) Moderate impact
(4) Significant impact
(5) Extreme impact
(6) I did not experience censorship outside the US
(7) I prefer not to answer

\noindent
6. In the US, the worst Internet censorship I experienced had
(1) No impact at all
(2) Slight impact
(3) Moderate impact
(4) Significant impact
(5) Extreme impact
(6) I did not experience censorship in the US
(7) prefer not to answer

\noindent
7. What is your age group? 

\noindent
8. What is your gender? 

\noindent
9. What is your occupation?

\noindent
10. We are planning follow-up interviews with similar questions. The interview will take about one hour and we will pay you 30 US dollars. If you are interested, please type your preferred contact information.

\section{Interview Guide}
\label{appendix:interview}

\begin{compactitem}
    \item Have you ever lived in a country where you experienced censorship? (a) What countries?
    
    \item $[$ For each country mentioned $]$ Please tell me about the censorship that you encountered in this country.
        
    \item Do you know or suspect who implemented and enforced the censorship you experienced in that country? 
    
    \item Have you ever evaded censorship in that country? Please tell me about any of your experiences with evading censorship in that country.
    
    \item What made you evade censorship? $[$ End for $]$

    \item $[$ If the participant experienced censorship in multiple countries $]$ Did you notice any similarities or differences between the censorship to which you were exposed in different countries?
   
    \item Were you ever exposed to false information in any of the countries where you lived? Which countries?

    \item $[$ For each country mentioned $]$ Do you remember any such incident that you would feel comfortable discussing with me?
    
    \item How did you know the information was false?
    
    \item Did you ever try to verify information that you encountered  in that country? Do you remember an example of information you tried to verify and how you verified it?
    
    \item Do you remember a case where you tried to verify information you encountered in that country, but found it difficult to do so? $[$ End for $]$

    \item $[$ If the participant was exposed to false information in multiple countries $]$ How do you think false information to which you were exposed in different countries compares?

    \item Have you heard of influence operations? Can you remember an example?

    \item Have you ever encountered information online that you suspected was promoted by an influence operation? Can you give me an example?
        
    \item Do you know who conducts these influence operations, such as any individuals or groups?

    \item Do you know how they operate and what they seek to achieve?

    \item Did censorship ever make it more difficult to verify information? Can you share an example?

    \item Have you ever evaded censorship to verify information you encountered online? Is there an example that you can share with me?
    
    \item Have you ever felt that censorship and false information are related in any other way? Can you share any examples from your personal experience?

    \item Have you ever felt that censorship and influence operations are related in any way? Can you share any examples from your personal experience?

\end{compactitem}

\end{document}